\documentclass[10 pt, journal]{IEEEtran}

\makeatletter
\def\ps@headings{%
	\def\@oddhead{\mbox{}\scriptsize\rightmark \hfil \thepage}%
	
	\def\@evenhead{\scriptsize\thepage \hfil \leftmark\mbox{}}%
	
	\def\@oddfoot{}%
	
	\def\@evenfoot{}}
	
\makeatother
\usepackage{cite}
\usepackage{multirow}
\usepackage{caption}
\usepackage{float}  
\usepackage{array}
\usepackage[colorlinks]{hyperref}

\usepackage{epsfig,bm,amsmath,amssymb,graphicx,setspace}
\usepackage[numbers,sort&compress]{natbib}
\usepackage{color,placeins}
\usepackage{multirow}
\usepackage{algorithm}
\usepackage{algpseudocode}
\usepackage{colortbl}
\usepackage{color}
\usepackage{dblfloatfix}
\usepackage{turnstile}
\usepackage{multicol}
\usepackage{array}

\usepackage{enumerate}
\usepackage{turnstile}
\usepackage{subcaption}
\usepackage{arydshln,paralist}
\usepackage{soul,tabularx,booktabs}
\usepackage{csquotes}
\usepackage{multirow}
\usepackage{adjustbox}
\usepackage{float}

\usepackage{comment} 
\newcommand{\inlinecomment}[1]{}

\usepackage{xcolor}
\usepackage{verbatim}
\usepackage[colorinlistoftodos]{todonotes} 
\usepackage{rotating}
\definecolor{usethiscolorhere}{rgb}{0.86666,0.78431,0.78431}
\usepackage{tikz}
\usetikzlibrary{mindmap}
\usetikzlibrary{calc,positioning}
\usepackage{lipsum,adjustbox}
\usetikzlibrary{decorations.pathmorphing}

\usepackage{pifont}

\makeatother

\usepackage{amsmath}

\begin{document}

\title{How Safe Is Your Data in Connected and Autonomous Cars: A Consumer Advantage or a Privacy \textcolor{black}{Nightmare?}}

\author{Amit Chougule, Vinay Chamola,~\IEEEmembership{Senior Member,~IEEE}, Norbert Herencsar, ~\IEEEmembership{Senior Member,~IEEE} and  \\ Fei Richard Yu,~\IEEEmembership{Fellow,~IEEE}

\thanks{Amit Chougule is with the Department of Electrical \& Electronics Engineering, BITS-Pilani, Pilani Campus, 333031, India. (e-mail: amitchougule121@gmail.com).}%

\thanks{Vinay Chamola is with the Department of Electrical and Electronics Engineering \& APPCAIR, BITS-Pilani, Pilani Campus, 333031, India. (e-mail: vinay.chamola@pilani.bits-pilani.ac.in).}%

\thanks{Norbert Herencsar is with the Department of Telecommunications, Faculty of Electrical Engineering
and Communication, Brno University of Technology, Technicka 3082/12, 61600 Brno, Czech Republic (email:
herencsn@vut.cz, herencsn@ieee.org).}%

\thanks{F. Richard Yu is with the Department of Systems and Computer Engineering, Carleton University, Ottawa, ON K1S 5B6, Canada. (e-mail: richard.yu@carleton.ca).}
}
\maketitle

\begin{abstract}
The rapid evolution of the automobile sector, driven by advancements in connected and autonomous vehicles (CAVs), has transformed how vehicles communicate, operate, and interact with their surroundings. Technologies such as Vehicle-to-Everything (V2X) communication enable autonomous cars to generate and exchange substantial amounts of data with real-world entities, enhancing safety, improving performance, and delivering personalized user experiences. However, this data-driven ecosystem introduces significant challenges, particularly concerning data privacy, security, and governance. The absence of transparency and comprehensive regulatory frameworks exacerbates issues of unauthorized data access, prolonged retention, and potential misuse, creating tension between consumer benefits and privacy risks. This review paper explores the multifaceted nature of data sharing in CAVs, analyzing its contributions to innovation and its associated vulnerabilities. It evaluates data-sharing mechanisms and communication technologies, highlights the benefits of data exchange across various use cases, examines privacy concerns and risks of data misuse, and critically reviews regulatory frameworks and their inadequacies in safeguarding user privacy. By providing a thorough analysis of the current state of data sharing in the automotive sector, the paper emphasizes the urgent need for robust policies and ethical data management practices. It calls for striking a balance between fostering technological advancements and ensuring secure, consumer-friendly solutions, paving the way for a trustworthy and innovative automotive future.
\end{abstract}



\section{Introduction}
\label{Section: Introduction}

\begin{figure*}[t]
\centering
\includegraphics[width=0.65\textwidth]{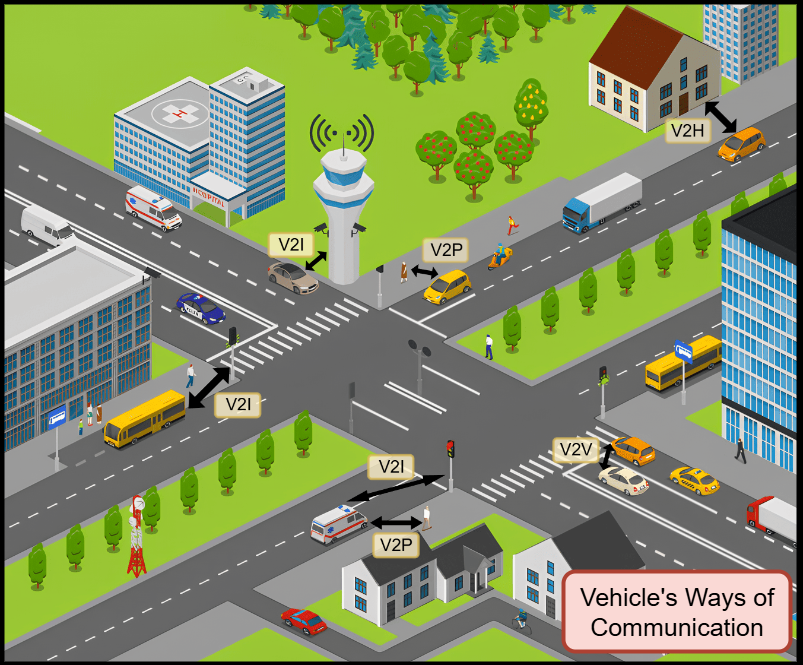}
    \hfill
    \caption{\textcolor{black}{Various ways a vehicle communicates during information and data sharing with real-world entities}}
    \label{fig:Vehicle's ways of communication}
\end{figure*}

The automobile sector is undergoing a significant transformation, marked by advancements over the past few years. Manufacturers are focused on improving engines, suspension systems, and overall vehicle designs to provide a more powerful and comfortable ride for both drivers and passengers. Engine improvements aim to enhance vehicle performance while maintaining fuel efficiency. Similarly, advancements in suspension technology contribute to a smoother and more comfortable journey for passengers.\inlinecomment{Additionally, innovations in vehicle design are centered around improving safety, with the goal of better protecting occupants in the event of a collision or accident.} Over time, the automobile sector has rapidly embraced and integrated advanced technologies and sensors, propelling vehicles to the next level of sophistication. By leveraging these innovations, vehicles now offer a wide array of features and services, significantly enhancing the user experience in terms of driving, safety, and comfort \cite{aledhari2023motion}. With the help of the Internet of Things (IoT), vehicles are capable of communicating with surrounding vehicles and real-world entities, enabling real-time services and features that further elevate the overall driving experience.

Vehicles can communicate and exchange data with real-world entities through various modes of communication, such as Vehicle-to-Everything (V2X), which is a broad category encompassing several sub-categories based on the entities involved and the use cases. In Vehicle-to-Vehicle (V2V) communication, vehicles wirelessly share critical information such as speed, location, direction, and other relevant data using protocols like Dedicated Short-Range Communication (DSRC) or Cellular. For communication with roadside units (RSUs), such as traffic lights, road signs, and toll booths, vehicles use Vehicle-to-Infrastructure (V2I) communication. Similarly, Vehicle-to-Pedestrian (V2P) communication enables vehicles to share data like pedestrian location, movement direction, speed, or vehicle information, enhancing safety for pedestrians. 
When a vehicle exchanges information with a broader network, such as cellular networks or the internet, Vehicle-to-Network (V2N) communication is employed. For data exchange with cloud-based services, Vehicle-to-Cloud (V2C) communication is used. In scenarios where vehicles communicate with personal devices like smartphones, wearables, or other IoT devices, Vehicle-to-Device (V2D) communication facilitates the exchange of data, typically relying on wireless technologies like Bluetooth, Wi-Fi, Near Field Communication (NFC), Ultra-Wideband (UWB), or 5G \cite{tang2023matching}. \inlinecomment{For electric vehicles (EVs), Vehicle-to-Home (V2H) communication allows the exchange of data or energy between the vehicle and the residential home. Additionally, Vehicle-to-Grid (V2G) communication enables EVs not only to charge from the grid but also to supply power back, contributing to grid stabilization, peak load management, and optimization of energy resources.}
Figure \ref{fig:Vehicle's ways of communication} showcases the various ways a vehicle communicates during information and data sharing with real-world entities.\par
\begin{table*}[t]
\centering
\footnotesize
\caption{Summary of Literature Review}  
\begin{tabular}{|c|m{2cm}|m{3.5cm}|m{5cm}|m{4.5cm}|}
\hline
\rowcolor[rgb]{0.949,0.949,0.949} \textbf{S No.} \vphantom{\rule{0pt}{3.0ex}} & \begin{center} \textbf{Author and Year} \end{center} & \begin{center} \textbf{Title of Paper} \end{center} & \begin{center} \textbf{Research Outcome} 
\end{center} & \begin{center} \textbf{Relevance in Current Work} \end{center} \\ 
\hline
\rowcolor[rgb]{0.902,0.925,0.949} \textbf{1.} & Ignatious \textit{et al.} (2022) \cite{ignatious2022overview} & An overview of sensors in Autonomous Vehicles & Systematic analysis of diverse sensors in AVs, emphasizing their role in decision-making and functional applications. & Highlights the significance of sensor and generated sensor data in autonomous vehicle systems \\ 
\hline
\textbf{2.} & He, Wenxue \textit{et al.} (2019) \cite{he2019overview} & Overview of V2V and V2I wireless communication for cooperative vehicle infrastructure systems & Comprehensive overview of V2V and V2I technologies, emphasizing their role in mitigating traffic congestion. & Role of V2V and V2I wireless communication technologies in AV and CAV applications \\ 
\hline
\rowcolor[rgb]{0.902,0.925,0.949} \textbf{3.} & Kabil, Ahmad \textit{et al.} (2022) \cite{kabil2022vehicle} & Vehicle to pedestrian systems: Survey, challenges and recent trends & Categorized V2P designs by VRU devices, technologies, and classifications, identifying trends and future directions. & Offers insights into safer interaction with pedestrians in AV contexts. \\ 
\hline
\textbf{4.} & Kerber, Wolfgang \textit{et al.} (2018) \cite{kerber2018data} & Data governance in connected cars: the problem of access to in-vehicle data & Discussed in-vehicle data access challenges using economic frameworks and governance strategies. & Informs policy design for data governance in connected vehicles. \\ 
\hline
\rowcolor[rgb]{0.902,0.925,0.949} \textbf{5.} & Taeihagh, Araz \textit{et al.} (2019) \cite{taeihagh2019governing} & Governing autonomous vehicles: emerging responses for safety, liability, privacy, cybersecurity, and industry risks & Reviewed technological risks and non-binding governmental responses to support AV development. & Provides insights into regulatory approaches for safe AV deployment. \\ 
\hline
\textbf{6.} & Fagnant, Daniel J \textit{et al.} (2015) \cite{fagnant2015preparing} & Preparing a nation for autonomous vehicles: opportunities, barriers and policy recommendations & Identified key impediments to AV adoption and proposed federal policy interventions. & Guides policy measures to enhance the adoption of AV technology. \\ 
\hline
\rowcolor[rgb]{0.902,0.925,0.949} \textbf{7.} & Collingwood, Lisa \textit{et al.} (2017) \cite{collingwood2017privacy} & Privacy implications and liability issues of autonomous vehicles & Provided an analysis of privacy and liability challenges in autonomous driving systems. & Highlights the importance of addressing privacy concerns in AV deployment. \\ 
\hline
\textcolor{black}{\textbf{8.}} & \textcolor{black}{Khan, Rabia, \textit{et al.} (2025)} \cite{khan2025decentralized} & \textcolor{black}{A Decentralized, Secure, and Reliable Vehicle Platoon Formation With Privacy Protection for Autonomous Vehicles} & 
\textcolor{black}{Proposed Privacy-Preserving Vehicle Platoon Formation Technique} & \textcolor{black}{Emphasizing the vital role of privacy protection in securing data and building trust in autonomous vehicles.} \\ 
\hline
\rowcolor[rgb]{0.902,0.925,0.949} \textcolor{black}{\textbf{9.}} & \textcolor{black}{Xiong, Jinbo \textit{et al.} (2024) \cite{khan2025decentralized}} &  \textcolor{black}{Privacy-Preserving Outsourcing Learning for Connected Autonomous Vehicles: 
Challenges, Solutions, and Perspectives} & \textcolor{black}{Identify the challenges, solutions, and future perspectives of privacy preservation in CAVs} & \textcolor{black}{Highlight the necessity of privacy preservation for CAVs}\\ 
\hline
\end{tabular}
\begin{center} 
\label{tab:literature_review}
\end{center}
\end{table*}
Connected and autonomous vehicles (CAVs), generate vast amounts of data while communicating with other vehicles and real-world entities. An individual CAV is estimated to generate over 300 TB (terabytes) of data annually \cite{tuxera2021,chougule2023comprehensive,grover2021edge,chamola2020fpga,verma2019cb}. This data includes information from various sensors, cameras, radar, and lidar systems, which provide real-time situational awareness, navigation updates, and interactions with external entities. When considering the aggregate data generated by all CAVs within a city, the volume becomes immense, further compounded by the frequency and types of data exchanged—ranging from traffic conditions and road status to V2V and V2I communication. This massive data flow is essential for enabling real-time decision-making, enhancing the driving experience, improving safety, and increasing the efficiency of transportation systems. However, managing, storing, and processing such large volumes of data presents significant challenges, particularly in terms of ensuring data privacy \cite{hassija2020blockchain, hassija2020secure}, security, and compliance with relevant regulations. \par

Currently, the data generated by CAVs is not adequately secured or managed to ensure the privacy of users or vehicle owners. Many companies, service providers, manufacturers, insurance companies, and maintenance organizations access this data to offer enhanced services to users. However, in numerous instances, these entities access the data without obtaining explicit permission from users, failing to inform them about the specific types of data being accessed, and retaining the data for longer periods than necessary. Such practices undermine user privacy and security, raising concerns about whether the accessibility of data genuinely benefits consumers or creates a privacy nightmare. The lack of proper data governance and transparency in data usage fosters a climate of uncertainty, where users are unaware of how their information is handled, who has access to it, and how long it is retained. This situation underscores the need for stronger regulatory frameworks, clearer guidelines, and more stringent data protection measures to ensure that users' privacy rights are respected and their data is used solely for legitimate, transparent purposes.\par

Considering these points, this review paper highlights the diverse benefits and risks associated with data sharing and data access within the automobile sector. It provides a comprehensive examination of the advantages that data exchange brings, such as improved vehicle performance, enhanced safety features, and the development of personalized services for users. Additionally, the paper discusses the risks inherent in the current system, particularly concerning privacy violations, unauthorized access to personal data, and the long-term retention of data beyond its useful purpose.

By analyzing both the positive and negative aspects, this paper aims to present a balanced perspective on how data sharing can drive innovation while addressing the pressing need for stronger regulatory frameworks and policies to protect users' privacy and security in the rapidly evolving automotive landscape. The main contributions of this review are as follows:
\begin{itemize}
\item An overview of the current state of data-sharing mechanisms in the real world for CAVs, highlighting the journey of data generation from sensors to its sharing with various real-world entities.
\item A review of the benefits of data sharing, with a focus on different use cases where sharing data enhances vehicle performance, safety, and user experience.
\item An examination of privacy concerns and risks associated with data sharing, addressing the potential for misuse of data by unauthorized entities.
\item A review of various regulatory frameworks and standards across different countries, analyzing the limitations of these frameworks in the automobile sector, particularly in the context of CAVs.
\end{itemize}

\section{Literature Survey}
\label{Section: Literature Survey}

Ignatious \textit{et al.} \cite{ignatious2022overview} systematically analyzes and synthesizes the roles played by diverse sensors utilized in autonomous vehicles (AVs), emphasizing their contributions to decision-making processes. The study further delves into the specific types of data these sensors capture, offering detailed insights into their functional applications in the AV ecosystem. Wenxue \textit{et al.} \cite{he2019overview} presents a comprehensive overview of wireless communication technologies, specifically V2V and V2I, with a focus on their practical applications. The review underscores their role in mitigating traffic congestion scenarios, providing a critical assessment of their use cases. Kabil, Ahmad \textit{et al.} \cite{kabil2022vehicle} conducts an extensive survey of various V2P system architectures, categorizing these designs based on critical criteria such as the type of Vulnerable Road User (VRU) device, underlying technologies, and VRU classifications (e.g., pedestrians, cyclists, or motorized two-wheelers). The analysis identifies emerging trends in V2P systems, offering a roadmap for future advancements in this domain. Kerber, Wolfgang \textit{et al.}  \cite{kerber2018data} explores the challenges of data governance in the context of connected vehicles, focusing specifically on the issue of in-vehicle data access. The study provides a nuanced discussion of policy debates surrounding these challenges and employs an economic framework, including market failure analysis, to dissect the implications of current governance strategies. Taeihagh, Araz \textit{et al.} \cite{taeihagh2019governing} reviews the technological risks inherent in autonomous driving and examines the evolving governmental responses to these challenges. The findings reveal that most governments have opted against implementing stringent regulatory measures to support AV development, favoring instead non-binding approaches such as the establishment of advisory councils and working groups. Fagnant, Daniel J \textit{et al.} \cite{fagnant2015preparing} investigates the barriers to large-scale adoption of autonomous vehicles, identifying key impediments to market penetration. The study proposes federal-level policy interventions aimed at facilitating a seamless transition to AV technology, including recommendations for the development of robust privacy standards. Collingwood, Lisa \textit{et al.} \cite{collingwood2017privacy} provides an in-depth analysis of the privacy and liability issues emerging in the context of autonomous driving. The discussion highlights the significant challenges posed by these issues, offering insights into the broader implications of this transformative technology. Table \ref{tab:literature_review} provides a summary of the literature review.
\section{Technological Overview}
\label{Section: Technological Overview}

\begin{table*}[!ht]
\centering
\footnotesize
\caption{\textcolor{black}{Sensors involved in self-driving cars}}  
\begin{tabular}{|c|m{3cm}|m{5cm}|c|m{4cm}|}
\hline
\rowcolor[rgb]{0.949,0.949,0.949} \textbf{No.} \vphantom{\rule{0pt}{3.0ex}} & {\centering \textbf{Sensor}} & {\centering \textbf{Function}}  & {\centering \textbf{Number of Sensors}} & {\centering \textbf{Position in Vehicle}} \\  
 \hline
\textbf{1.}  & LiDAR (Light Detection and Ranging) & Uses laser pulses to create a 3D map of the environment by measuring distances to objects \cite{kamalkumar2022lidar, abbasi2022lidar}. & \textcolor{black}{1-2+} & Mounted on the roof or front bumper \\ 
\hline
\rowcolor[rgb]{0.902,0.925,0.949}  \textbf{2.}  & Radar & Uses radio waves to detect objects and measure their speed, distance, and angle relative to the car \cite{hussain2022drivable, hussain2020multiple, liu2021robust}. & 4-6 & Front and rear bumpers, sides \\ 
\hline
\textbf{3.}  & Monocular Camera & A single-lens camera used for capturing visual data, such as detecting objects, lane markings, and signs \cite{rashed2021bev, zhao2020dynamic}. & 1-2 & Windshield or front grille \\ 
\hline
\rowcolor[rgb]{0.902,0.925,0.949}  \textbf{4.}  & Stereo Camera & Uses two cameras spaced apart to perceive depth, helping in detecting the distance and size of objects \cite{zaarane2020distance}. & 1 & Front of the vehicle (windshield or grille) \\ 
\hline
\textbf{5.}  & Thermal Camera & Detects heat signatures from objects, useful for night driving and detecting pedestrians or animals  \cite{bhadoriya2022vehicle, shin2023deep}. & 1 & Front grille \\ 
\hline
\rowcolor[rgb]{0.902,0.925,0.949}  \textbf{6.}  & Wide-Angle Camera & Captures a broader view of the surroundings, helpful in detecting objects in the peripheral view \cite{kumar2023surround}. & 2-4 & Front, rear, and sides \\ 
\hline
\textbf{7.}  & Ultrasonic Sensors & Used for short-range object detection, typically for parking and low-speed maneuvers \cite{xu2018analyzing}. & 8-12 & Front and rear bumpers, side mirrors \\ 
\hline
\rowcolor[rgb]{0.902,0.925,0.949}  \textbf{8.}  & \textcolor{black}{Global Positioning System (GPS)} & Provides the car’s precise location and assists in navigation \cite{wong2020mapping}. & 1 & Inside the vehicle, connected to the dashboard \\ 
\hline
\textbf{9.} & Odometry Sensors & Track wheel movement to estimate the car’s position over time \cite{brunker2018odometry, mohamed2019survey}. & 4 (one per wheel) & Attached to each wheel \\ 
\hline
\rowcolor[rgb]{0.902,0.925,0.949} \textbf{10.} & Infrared Sensors & detect heat signatures from objects, useful for night driving and detecting pedestrians \cite{thakur2018infrared}. & 2-4 & Front and rear bumpers \\ 
\hline
\textbf{11.} & Temperature Sensors & Monitor the temperature of various car components to prevent overheating and ensure optimal performance \cite{dong2024sensors}. & 2-3 & Engine compartment, battery, and cabin \\ 
\hline
\rowcolor[rgb]{0.902,0.925,0.949}  \textbf{12.} & Gyroscope & Measures the car’s orientation and helps maintain balance and control \cite{nazemipour2020mems}. & 1 & Inside the vehicle \\ 
\hline
\textbf{13.} & Accelerometers & Detect changes in speed and direction to assist with navigation and safety features \cite{zylius2017investigation}. & 1-2 & Inside the vehicle, integrated with IMU or centrally located \\ 
\hline
\rowcolor[rgb]{0.902,0.925,0.949} \textbf{14.} & Proximity Sensors & Detect nearby objects to prevent collisions and assist in low-speed navigation \cite{braun2015capseat}. & 6-8 & Front, rear, and sides \\ 
\hline
\end{tabular}
\begin{center} 
\label{tab:self_driving_sensors}
\end{center}
\end{table*}

CAVs rely on a complex network of advanced technologies to operate safely and efficiently. Central to these vehicles are an array of sophisticated sensors that work together to perceive and interpret the environment. These sensors include LiDAR, Radar, monocular cameras, stereo cameras, thermal cameras, wide-angle cameras, ultrasonic sensors, GPS, odometry sensors, infrared sensors, temperature sensors, gyroscopes, accelerometers, and proximity sensors. 
\inlinecomment{Figure \ref{fig:Various devices and sensors in self-driving and connected cars}  represents various devices and sensors utilized in CAVs. }These sensors generate vast amounts of data, which is crucial for enabling real-time decision-making, improving vehicle safety, and enhancing the overall driving experience. The data also plays a key role in advancing CAVs ecosystem.\par
Various sensors generate different types of data in diverse formats, contributing to a comprehensive understanding of an environment or system. Monocular cameras produce 2D image frames or video streams, where the data is captured as pixel values in formats like JPEG, PNG, or video formats \cite{rashed2021bev}. These cameras typically generate 30-60 frames per second (FPS), capturing visual information in RGB or grayscale. Similarly, stereo cameras generate synchronized 2D image pairs that can be processed into disparity maps for depth estimation, also producing 30-60 FPS \cite{zaarane2020distance}. \inlinecomment{The thermal camera, while similar in function to monocular cameras, captures temperature distribution as pixel data where each pixel represents a specific temperature in degrees Celsius or Fahrenheit, often in grayscale or color-mapped formats like TIFF or PNG \cite{bhadoriya2022vehicle, shin2023deep}.}Wide-angle cameras extend the field of view while capturing similar pixel-based image or video data, also generating 30-60 FPS \cite{kumar2023surround}. \par 
Moving on to point cloud data, LiDAR (Light Detection and Ranging) sensors generate 3D point clouds, which represent the geometry of the environment through X, Y, Z coordinates and intensity values, typically stored in formats such as \textcolor{black}{Point Cloud Data (PCD)} or LAS \cite{kamalkumar2022lidar, abbasi2022lidar}. These sensors typically produce 10-20 point cloud updates per second, with high-end models generating up to 100 point clouds per second. Radar sensors also produce point cloud data but focus on detecting objects’ distance, velocity, and occasionally their size and shape \cite{hussain2022drivable, hussain2020multiple}. Radar data points typically include range in meters, velocity in meters per second, and sometimes angle in degrees along with reflection intensity, with data generated 10-20 times per second. For range and proximity measurements, ultrasonic sensors provide distance measurements to the nearest object, with data expressed in meters or centimeters and generated 10-50 times per second \cite{xu2018analyzing}. \inlinecomment{Infrared sensors generate distance measurements or proximity detection data, which may be recorded as distance in meters or binary data indicating the presence or absence of an object, typically generating data 10-60 times per second \cite{thakur2018infrared}.}Proximity sensors, in general, produce binary data or distance measurements, outputting either a simple on/off signal or a measured distance, typically in meters or centimeters, and generating data 10-100 times per second \cite{braun2015capseat}. \par
In the realm of position and orientation, \textcolor{black}{Global Positioning System (GPS)} sensors output geospatial coordinates in terms of latitude, longitude, and altitude, along with time, typically measured in degrees and meters, and updated 1-10 times per second \cite{wong2020mapping}. Odometry sensors track relative position changes, velocity, and occasionally orientation, outputting data in meters for distance, meters per second for velocity, and degrees for orientation, usually generating data 10-100 times per second \cite{brunker2018odometry, mohamed2019survey}. Gyroscopes measure angular velocity across three axes, generating data in degrees per second or radians per second, with high operational rates of 50 to 1000 data updates per second \cite{nazemipour2020mems}. Accelerometers, which measure linear acceleration along three axes, output data in meters per second squared (m/s²) and generate data at similar rates \cite{zylius2017investigation}. \inlinecomment{Lastly, temperature sensors capture ambient or object temperature data, outputting values typically in degrees Celsius (°C) or Fahrenheit (°F), with data updates typically occurring 1-10 times per second \cite{dong2024sensors}.} These sensors collectively enable a detailed and multi-dimensional understanding of both static and dynamic aspects of an environment or system. Table \ref{tab:self_driving_sensors} lists various sensors involved in CAVs. \inlinecomment{Also, Figure \ref{fig:CAV Data Communication Layers} illustrates the CAV Data Communication Layers, emphasizing the structured flow of data in CAV's. These layers include Physical layer, Data Link layer, Network layer, Transport layer, and Application layer.}

\section{Benefits of Data Sharing}
\label{Section: Benefits of Data Sharing}

\begin{figure*}[t]
\centering
\includegraphics[width=0.65\textwidth]{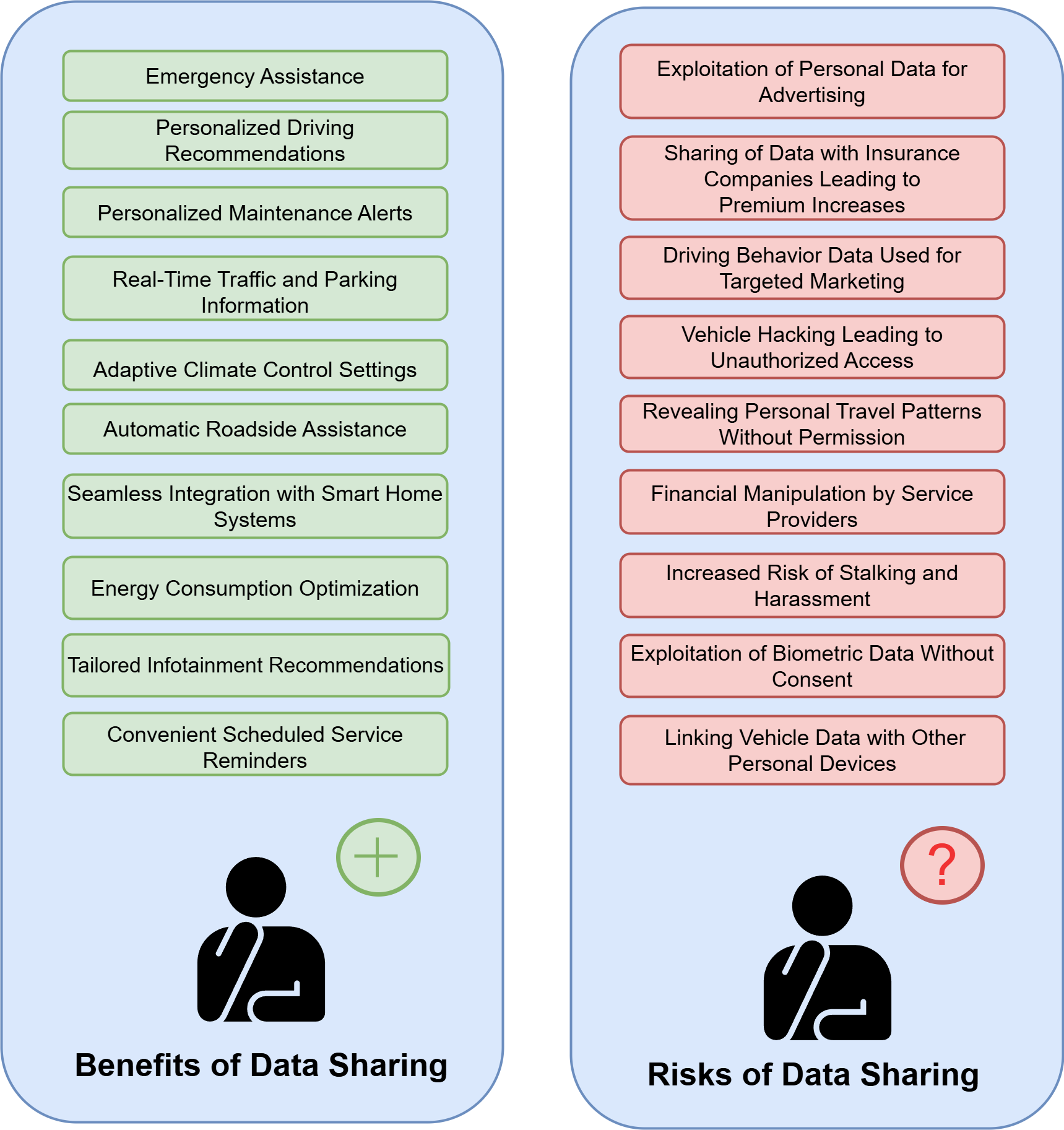}
    \hfill
    \caption{Benefits and Risks of Data Sharing in automobile sector}
    \label{fig: Benefits and Risks of Data Sharing}
\end{figure*}

As CAVs become more embedded in daily life, they are transforming the interaction between consumers and vehicles, shifting them from mere modes of transportation to intelligent systems that anticipate and respond to individual needs. The integration of data-sharing capabilities in CAVs with real-world entities has unlocked a myriad of benefits for consumers, enhancing convenience, safety, and the overall driving experience. By allowing vehicles to collect, process, and share data, manufacturers and service providers can offer highly personalized and responsive features that were previously unimaginable. Continuous data exchange between the vehicle, its surroundings, and various service providers enables unprecedented levels of safety, convenience, and personalization. \par
The benefits of data sharing in CAVs extend beyond basic functionality, providing consumers with a more comfortable and streamlined driving experience. Automating many aspects of vehicle management and interaction, data sharing reduces the need for drivers to manually handle tasks like booking services or managing maintenance schedules \cite{prytz2015predicting}. This automation saves time and simplifies life by delivering services exactly when and where they are needed, often without the driver even realizing it \cite{wang2021restaurant,shambour2023restaurant}. Additionally, data sharing enhances safety by enabling vehicles to react to their environment in real-time, offering a level of protection and reassurance that is difficult to achieve otherwise. Furthermore, with data-driven insights, vehicles can offer more choices and options tailored to individual preferences, giving drivers greater flexibility and control over their driving experience. This seamless integration of services and customization makes driving more enjoyable and efficient, fundamentally enhancing the value of CAVs for modern consumers. This section explores the various benefits that data sharing brings to consumers, demonstrating how it can significantly enhance their driving experience and lifestyle. Figure \ref{fig: Benefits and Risks of Data Sharing} illustrates the multifaceted benefits and risks associated with data sharing in CAVs.

\subsection{Emergency Assistance} When a vehicle equipped with autonomous driving technology encounters an emergency situation, such as an accident or sudden medical emergency, it can leverage a range of sensor data and connectivity features to provide timely and effective assistance \cite{nguyen2023risk, chougule2023comprehensive, zanella2014internet}. Typically, this involves the vehicle’s sensors—such as cameras, radar, lidar, and GPS—detecting anomalies or sudden changes in driving conditions. For instance, if the vehicle detects a collision or an abrupt stop, it can automatically relay critical information, including the exact location, the severity of the incident, and the vehicle’s condition, to emergency services and nearby vehicles. This data is shared through V2V and V2I communication channels, which ensure that first responders are alerted and can reach the scene quickly \cite{liu2022towards}. Moreover, the data can include real-time diagnostic information, such as the status of airbags, the integrity of safety systems, and even the health status of occupants if medical monitoring systems are in place. This enables emergency responders to prepare adequately before arriving at the scene, potentially saving lives and reducing the impact of the incident. For consumers, this integration of emergency assistance provides enhanced safety and prompt help. It ensures that in critical situations, help is not only prompt but also well-informed, leading to more effective emergency response and improved overall safety on the roads.

\subsection{Personalized Driving Recommendations} Personalized driving recommendations represent a significant benefit derived from the sharing of sensor data and connected car information \cite{mantouka2022deep}. This process involves utilizing a vehicle’s data—such as driving patterns, speed, braking habits, and navigation preferences—to offer tailored suggestions that enhance the driving experience and improve safety \cite{liao2024review}. As a vehicle collects data from its various sensors and systems, it accumulates detailed insights into the driver’s habits and preferences. For example, if the vehicle’s sensors detect frequent hard braking or aggressive acceleration, the system can analyze these behaviors to recommend more fuel-efficient driving practices or suggest maintenance checks if they indicate underlying issues \cite{liao2024review}. Additionally, the system can use real-time data to provide contextual recommendations. For instance, if the vehicle detects that the driver frequently travels on certain types of roads or in specific weather conditions, it can suggest optimal routes \cite{rogers1998personalized, ge2019route}, adjust driving modes, or recommend nearby points of interest such as gas stations or restaurants based on the driver’s preferences and habits. These recommendations are often delivered through the vehicle’s infotainment system, offering a user-friendly interface for drivers to act upon the suggestions. By personalizing these insights, drivers can improve their driving efficiency, reduce costs related to fuel and maintenance, and enhance their overall driving comfort \cite{gilman2015personalised, rios2019fuel}. For consumers, personalized driving recommendations translate into a more enjoyable and efficient driving experience, with tailored advice that helps to optimize vehicle performance and safety based on individual driving patterns and preferences.

\subsection{Personalized Maintenance Alerts}
Personalized maintenance alerts are a valuable feature enabled by the sharing of sensor data and connected car information \cite{solanki2017iot}. This system leverages data collected from a vehicle’s sensors, diagnostic tools, and connectivity modules to provide timely and tailored notifications regarding vehicle maintenance needs \cite{bickelhaupt2024towards,denton2020advanced}. Each vehicle continuously monitors its condition through various sensors that track parameters such as engine performance, brake wear, tire pressure, and fluid levels. This real-time data is analyzed to detect deviations from normal operating conditions or signs of potential issues. For instance, if the system detects that the brake pads are nearing the end of their life or that the engine oil level is low, it generates personalized maintenance alerts \cite{rognvaldsson2018self}. These alerts are communicated to the vehicle owner through the infotainment system. The information is customized based on the specific condition of the vehicle and its usage patterns, ensuring that the alerts are relevant and actionable. For example, a vehicle that frequently drives in harsh conditions might receive earlier warnings about maintenance needs compared to a vehicle used under more typical conditions. By providing these personalized maintenance alerts, the system helps drivers stay informed about their vehicle’s health and avoid potential breakdowns \cite{jeong2018integrated}. Timely maintenance not only enhances vehicle performance and safety but also helps to prevent costly repairs by addressing issues before they escalate \cite{jeong2018integrated}. For consumers, personalized maintenance alerts translate into greater convenience and peace of mind, ensuring that maintenance tasks are not overlooked and contributing to the vehicle’s longevity and optimal performance while reducing the risk of unexpected failures and associated repair costs.

\subsection{Real-Time Traffic and Parking Information}
Real-time traffic and parking information is a significant advantage of sharing sensor data and CAVs technology. This system leverages data from a vehicle’s sensors, GPS, and connectivity features to provide drivers with current, actionable information that improves their driving experience \cite{shen2012real,chen2020short,wang2020truck}. CAVs gather and transmit data about traffic conditions, including congestion, road closures, and accidents. This data, sourced from other CAVs, traffic management systems, and real-time traffic services, is aggregated and analyzed to offer dynamic route recommendations \cite{azimjonov2021real,chougule2023novel, raja2024intuitive, chen2024empowering}. Drivers receive updates that help them avoid traffic jams and minimize travel time by suggesting alternative routes when necessary. In addition to traffic information, CAVs also deliver real-time parking updates. Through data from parking sensors, infrastructure communication, and connected parking facilities, the system informs drivers about available parking spaces, their locations, and associated costs \cite{gahlan2016gps,tripathi2020smart,chamola2024overtaking}. This functionality is particularly valuable in urban areas where parking can be scarce and difficult to find \cite{parmar2020study}. By integrating this information into the vehicle’s navigation system or a connected app, drivers can make informed decisions on the road. For instance, if heavy traffic is detected on the planned route, the system can suggest a more efficient alternative. Likewise, if parking is available near the destination, it provides directions to the nearest spot. For consumers, these real-time updates enhance convenience and efficiency, alleviating stress related to traffic delays and parking difficulties. This results in a more streamlined and enjoyable driving experience, saving time and reducing frustration.

\subsection{Adaptive Climate Control Settings}
Adaptive climate control settings represent a sophisticated application of shared sensor data and CAVs technology to enhance driver comfort and vehicle efficiency \cite{wang2018model}. This system utilizes data from a vehicle’s internal and external sensors to automatically adjust climate control settings based on real-time conditions and individual preferences. Sensors within the vehicle monitor factors such as cabin temperature, humidity, sunlight intensity, and even the driver’s and passengers’ preferences \cite{daly2011automotive}. External data, including current weather conditions and the vehicle’s environmental context, is also integrated \cite{chakrabarty2013analysis}. For instance, if the sensors detect a rapid increase in outside temperature, the system can preemptively adjust the air conditioning to maintain a comfortable cabin environment \cite{adomah2022safety,ahmed2018driver}. Additionally, adaptive climate control leverages data on the driver’s behavior and historical preferences to refine its adjustments. If the system observes that the driver prefers a cooler setting at certain times of the day or under specific conditions, it adjusts climate control to match these preferences automatically. Integration with the vehicle’s connected infrastructure allows for continuous adaptation. For example, if the vehicle’s sensors detect that the driver is entering a sunny area or approaching a parking garage, the climate control can adjust to ensure comfort upon arrival. For consumers, adaptive climate control settings offer significant benefits in convenience and comfort. By automatically adjusting to changing conditions and personal preferences, the system enhances the driving experience, reduces the need for manual adjustments, and contributes to overall energy efficiency by optimizing climate control based on real-time data.

\subsection{Automatic Roadside Assistance}
Automatic roadside assistance is a critical feature enabled by the integration of sensor data and connected car technology \cite{kadas2013role}. This system leverages real-time data from a vehicle’s sensors and connectivity features to provide prompt and efficient assistance in case of breakdowns or other roadside emergencies. When a vehicle encounters an issue such as a flat tire, engine malfunction, or low fuel, onboard sensors detect the problem and automatically trigger an alert \cite{chand2020analysis}. This alert includes detailed information about the vehicle’s condition, location, and the nature of the issue. The data is transmitted to a central roadside assistance service or directly to emergency response teams, depending on the vehicle's connectivity setup. For example, if the vehicle’s diagnostic system identifies a critical engine problem, it can send a signal to a roadside assistance provider or the nearest vehicle manufacturer’s showroom with information on the vehicle’s exact location and the detected issue. This allows the provider to dispatch the appropriate help, whether it’s a tow truck, a repair technician, or a service vehicle equipped with the necessary tools and parts to address the problem. The system can also use data from the vehicle’s GPS to provide precise location information, reducing the time it takes for assistance to arrive \cite{goregaonkar2013safe, wang2020truck}. Additionally, in some cases, the vehicle might offer basic troubleshooting advice or self-help options. For consumers, automatic roadside assistance offers significant peace of mind. It ensures that help is dispatched quickly and efficiently when needed, minimizing the time spent stranded and reducing the stress associated with vehicle breakdowns. This feature not only enhances safety but also contributes to a more reliable and stress-free driving experience.

\subsection{Seamless Integration with Smart Home Systems}
Seamless integration with smart home systems represents a sophisticated use of connected car technology to enhance the convenience and functionality of both the vehicle and the home environment. This integration allows vehicles to communicate with smart home devices, creating a cohesive and intuitive experience for users. When a vehicle is equipped with smart connectivity features, it can interface with various smart home systems such as heating or air conditioning, lighting, security systems, and even appliances \cite{yaici2022internet, li2017smart}. For example, as a driver approaches their home, the vehicle’s GPS and proximity sensors can trigger a series of pre-configured actions. The vehicle might send a signal to the home’s smart heater or air conditioning to adjust the temperature to a preferred setting, or it could activate outdoor lighting to ensure a well-lit entrance. Additionally, smart home integration allows for remote control of home systems through the vehicle’s infotainment system \cite{deserno2020transforming}. If the driver needs to adjust the home’s heating or cooling while still on the road, they can do so via the vehicle’s interface \cite{yaici2022internet}. This functionality can be particularly useful for ensuring that the home environment is comfortable and ready upon arrival. Security and convenience are also enhanced through this integration. For instance, if the vehicle’s sensors detect that the driver is approaching home and the garage door is not yet open, the system can automatically trigger the garage door opener \cite{torad2022smart}. Similarly, if the vehicle’s system detects that the driver has left the house and the security system was not activated, it can send a notification or even activate the security system remotely \cite{chen2016multi}. For consumers, seamless integration with smart home systems provides a higher level of convenience, comfort, and security. It simplifies the management of both vehicle and home systems, ensuring that users can effortlessly coordinate their environments and enjoy a more connected, efficient lifestyle.

\subsection{Energy Consumption Optimization}
Energy consumption optimization in CAVs extends beyond managing energy use during travel to include proactive measures for battery or fuel management. When it comes to charging electric vehicles or refueling hybrid vehicles, the system utilizes real-time data to enhance convenience and efficiency. The vehicle's sensors continuously monitor battery levels, driving patterns, and energy consumption, sharing this data with a network of connected services \cite{tseng2017personalized}. When the battery or fuel reaches a low threshold, the system can automatically identify nearby charging stations or fuel stations. Using real-time data on station availability and operational status, it can reserve a spot or schedule an appointment, ensuring that drivers can efficiently manage their energy needs without delays. Moreover, the system can communicate with other vehicles and infrastructure to gather information on the availability and status of nearby stations \cite{tian2016real, lin2024pdqn}. This shared data helps in optimizing the route to include stops at stations with the best availability and pricing, reducing the time spent searching for a suitable location. For consumers, this integration simplifies energy management by automating the process of finding and reserving charging or refueling spots. It minimizes disruptions and ensures that energy needs are met efficiently, contributing to a smoother and more convenient driving experience.

\subsection{Tailored Infotainment Recommendations}
Tailored infotainment recommendations leverage connected car technology and sensor data to provide a highly personalized in-vehicle entertainment and information experience. This system utilizes data about the driver’s preferences, driving patterns, and real-time contextual information to suggest content that enhances the driving experience. The vehicle’s infotainment system collects and analyzes data such as the driver’s music preferences \cite{ccano2017mood}, favorite radio stations, podcast subscriptions, and frequently visited destinations. By integrating this data with real-time inputs, such as current traffic conditions or the time of day, the system can offer recommendations tailored to the driver’s immediate context \cite{raja2024intuitive}. For example, if the vehicle detects that the driver is on a long journey, it might suggest engaging podcasts or audiobooks based on previous listening habits. Alternatively, during a daily commute, the system could recommend news updates or favorite radio stations to match the time of day or driving duration \cite{ccano2017mood}. Additionally, the system can use location-based data to suggest nearby points of interest, such as restaurants or attractions, based on the driver’s preferences and past behavior \cite{wang2021restaurant,shambour2023restaurant}. If the vehicle is equipped with voice recognition, drivers can interact with the infotainment system to request specific types of content or make adjustments based on real-time feedback. For consumers, tailored infotainment recommendations enhance the driving experience by providing personalized and relevant content that aligns with individual preferences and driving conditions. This not only makes travel more enjoyable but also ensures that entertainment and information are seamlessly integrated into the driving experience, contributing to a more engaging and convenient journey.

\subsection{Convenient Scheduled Service Reminders}
Convenient scheduled service reminders extend beyond vehicle maintenance to include various other vehicle-related services and functions \cite{prytz2015predicting}. Through the integration of connected car technology and real-time data, these reminders enhance overall vehicle management and driver convenience. For instance, the system can automatically notify drivers about upcoming registration renewals or insurance policy expirations based on data inputs and historical records. By keeping track of these important deadlines, the system helps ensure compliance with legal requirements and avoids potential fines or lapses in coverage. In addition to compliance-related reminders, the system can also provide alerts for routine checks or updates that enhance vehicle functionality. Furthermore, the system can alert drivers to seasonal tasks, such as preparing the vehicle for winter conditions \cite{mills2019changing,malmivuo2017effects} or summer road trips \cite{liu2022impact}. For example, it might remind drivers to check tire pressure \cite{liu2022impact}, replace wiper blades, or verify fluid levels before embarking on long journeys, and it also checks the vehicle servicing slots with the nearest service point by sharing information with the service points. For consumers, these scheduled service reminders offer a streamlined approach to managing various aspects of vehicle ownership. By automating and personalizing reminders for a range of services and requirements, the system ensures that critical tasks are completed on time, enhancing vehicle performance, safety, and overall convenience.

\section{Privacy Concerns and Risks}
\label{Section: Privacy Concerns and Risks}

\textcolor{black}{The integration of data-sharing capabilities in CAVs introduces substantial privacy challenges despite its benefits in convenience, safety, and personalization. As vehicles evolve to become more intelligent and interconnected, the scale and sensitivity of collected and shared data expand rapidly, encompassing not only location and driving behavior but also personal preferences, biometric data, and interactions with external systems. This heightened connectivity, while facilitating a seamless and customized driving experience, simultaneously increases susceptibility to data breaches, unauthorized access, and surveillance. The reliance on continuous data exchange intensifies concerns regarding data control, storage practices, and usage purposes. In the absence of strong privacy protections and clearly defined regulatory frameworks, consumer data remains exposed to potential misuse, eroding the trust essential for the widespread acceptance of CAV technologies. \inlinecomment{Figure \ref{fig: Risks associated with data sharing in CAVs} illustrates various scenarios highlighting the risks linked to data sharing in CAV contexts.}This section presents an analysis of the primary privacy concerns and risks stemming from data-sharing practices in CAVs.}

\subsection{Exploitation of Personal Data for Advertising}
When CAVs data is shared, one significant risk involves the exploitation of personal data for advertising purposes \cite{chiasserini2017advertisement, wang2019efficiently,einziger2018scheduling}. In connected vehicles, various sensors collect a broad array of data, including GPS locations, driving habits, and in-car interactions. When shared with third parties, this data can be used to build highly detailed consumer profiles, enabling advertisers to target individuals with highly personalized ads based on their driving patterns and preferences. For instance, if a vehicle's GPS data shows frequent visits to specific types of businesses or geographical locations, advertisers may use this information to tailor marketing campaigns directly to the vehicle owner’s habits and preferences \cite{abrougui2010location, pradipta2011profiling}. While this personalization can enhance ad relevance, it also raises significant privacy concerns. Consumers may find their personal choices and behaviors monitored and utilized without their explicit consent or full awareness. Additionally, the aggregation of data from multiple sources can lead to even more detailed profiling, where consumer behavior is tracked in real-time and analyzed to predict future actions \cite{wang2023data}. Such practices can result in invasive advertising strategies, with consumers inundated by highly targeted ads, potentially leading to an erosion of privacy and a sense of constant surveillance. In short, while sharing CAVs data can enable advanced advertising techniques, it also poses considerable risks to consumer privacy. The detailed nature of the data collected and its use in personalized advertising can create an environment where consumers feel their personal space and choices are being exploited, leading to potential mistrust and discomfort.

\subsection{Sharing of Data with Insurance Companies Leading to Premium Increases}
Data from connected vehicles and recent on-road vehicles equipped with various sensors and connectivity features is often shared with insurance companies via vehicle-to-insurer communication channels \cite{baecke2017value}. This data is shared to track the vehicle’s condition, schedule upcoming maintenance (such as oil changes or services based on kilometers driven), and ensure that insurance policies remain active. However, insurance companies may also gain indirect access to this data through third-party aggregators or unauthorized agreements with automakers, often without the driver’s explicit consent or full knowledge.

The data collected includes detailed driving behaviors, such as speed, braking patterns, location data, and environmental conditions. Insurance companies use this information to refine risk assessment models, categorizing drivers into different risk tiers based on their driving habits \cite{arumugam2019survey}. For instance, frequent hard braking might lead to a driver being classified as high-risk, even if such behavior is contextually justified, such as reacting to sudden traffic changes \cite{arumugam2019survey}. This can result in increased premiums for drivers, often based on isolated driving incidents, creating financial stress for consumers who otherwise drive safely. The context of driving behavior is often overlooked, leading to unfair classifications and unpredictable premium hikes \cite{bian2018good}. Moreover, insurers might not restrict their use of this data to risk assessments alone. They could misuse the information for more invasive purposes, such as selling location and behavioral data to third parties or advertisers. Location data, in particular, is highly valuable to companies in various sectors, and if insurers engage in selling this information, consumers could be subjected to unwanted and highly personalized ads. This practice erodes privacy and commodifies personal driving data.

The broader implications for consumer rights and privacy are significant \cite{van2022informational}. The distinction between voluntary data sharing and indirect exploitation becomes increasingly blurred, especially when drivers are unaware of the extent of data being shared and how it is used. Many may not realize their insurance premiums are influenced by data collected without their direct consent, creating a lack of transparency in the industry \cite{duri2004data}. This lack of clarity can lead to consumer distrust and raise legal and ethical concerns about personal data use in some jurisdictions. While data collection can improve the precision of insurance risk assessments, it also opens the door to misuse, unfair treatment, and consumer distrust. The exploitation of driving data, often obtained without full consent, leaves consumers vulnerable both financially and in terms of personal privacy.

\subsection{Driving Behavior Data Used for Targeted Marketing}
Sensor data from modern on-road vehicles is increasingly utilized by marketing companies through partnerships between automakers and third-party data brokers. This data, including GPS location history, driving patterns such as speed and acceleration, and in-car preferences like entertainment choices and frequently visited destinations, is often accessed indirectly. Such access is facilitated through agreements embedded in lengthy terms of service or vague privacy policies, frequently without explicit user consent.

Marketing firms leverage this rich dataset to construct detailed consumer profiles for targeted advertising. For example, frequent visits to specific types of retail locations, such as shopping malls or fast-food chains, enable marketers to tailor ads that resonate with those habits \cite{ purwanto2022customer}. GPS data further allows for geographically targeted marketing, sending promotions for nearby businesses to the driver's phone or in-car system when they are in proximity to those establishments \cite{bauer2016location}. Driving behavior data also informs more nuanced marketing strategies. Patterns like frequent long road trips may lead to targeted ads for food pickups, hotels, travel packages, or vehicle maintenance services. Similarly, in-car preferences, such as music choices or app usage, can result in recommendations for related media content or subscription services.

While this personalized advertising can enhance ad relevance, it raises significant privacy concerns \cite{tong2020personalized, toch2012personalization}. The depth of data available to marketers can feel intrusive, with consumers potentially being monitored and profiled based on their driving habits. Unlike traditional advertising, which relies on broader demographic data, this approach offers a level of personalization that may be unsettling to some drivers, who might not fully realize the extent of their data being used. There is also a risk of misuse. For example, drivers who frequently visit healthcare facilities might receive ads for medical services or insurance, which could be uncomfortable if they prefer not to commercialize this aspect of their lives \cite{bauer2016location}. Additionally, data about driving routes revealing personal habits—such as visits to religious institutions or gyms—could lead to ads that are perceived as intrusive or inappropriate. The risk of data breaches or unauthorized sales of personal information further compounds these issues. Inadequately protected data could be exposed in breaches or sold to unscrupulous third parties, leading to potentially invasive marketing campaigns or identity theft. Moreover, the lack of transparency in data sharing practices adds to the concern. Consumers may believe they are sharing limited data for essential vehicle functions, while a broader range of personal information is actually being transmitted to marketing firms. This lack of clarity can erode trust, as drivers feel they have limited control over their data and its use. In short, while data sharing enhances targeted marketing precision, it introduces significant privacy and ethical challenges. The ability to create highly personalized ads based on detailed driving patterns and location data can feel invasive and contribute to a sense of constant surveillance. Additionally, the indirect and often unauthorized access to this data by third parties undermines consumer trust and exposes drivers to potential misuse, from invasive advertising to severe security risks.

\subsection{Vehicle Hacking Leading to Unauthorized Access}
Modern on-road vehicles are equipped with a variety of connectivity and communication technologies, such as Wi-Fi, Bluetooth, and cellular networks, enabling seamless interaction with other vehicles and real-world entities \cite{tsugawa2002inter, demba2018vehicle, alladi2020consumer}. These systems support useful operations like navigation, traffic management, and automated payments, enhancing convenience for drivers. However, these same technologies also introduce vulnerabilities that can be exploited by cybercriminals. Unauthorized access to a vehicle’s sensitive data—such as location history, driving behavior, or personal information like in-car payment methods—can occur without the driver’s knowledge or consent. Once hackers breach a vehicle’s systems, the risks extend far beyond privacy concerns. Remote control of critical functions, including braking, steering, and acceleration, becomes a possibility, directly endangering the safety of the driver and passengers \cite{ring2015connected,jafarnejad2015car}. In more extreme cases, hackers may disable essential safety features like airbags or anti-lock brakes, significantly increasing the risk of accidents. The ramifications of such exploits can be catastrophic, potentially leading to life-threatening situations.

Beyond safety risks, cybercriminals can misuse the stolen data for financial gain. Personal details, including addresses and financial information linked to vehicle payment systems, can be sold on the dark web or used for identity theft. This compromises drivers' financial security, and stolen location data may lead to more serious crimes, such as burglaries when hackers determine the driver is away from home. Hackers can also track vehicles in real-time, enabling them to facilitate stalking, physical harm, or robbery \cite{ring2015connected,jafarnejad2015car}. The ability to track vehicle movements has broader implications, such as industrial espionage, where sensitive business information or the movements of high-profile individuals are exploited and sold. Such incidents further erode consumer trust in connected car technologies, as drivers expect these systems to be secure. With the growing integration of connectivity in everyday driving, concerns about cybersecurity become more prominent. Weak security measures make connected vehicles appealing targets for skilled hackers, undermining the safety and integrity of these advanced systems.

\textcolor{black}{Additionally, advanced AI-driven cyberattacks pose a significant threat to consumer privacy by enabling automated hacking techniques capable of infiltrating vehicle networks and connected services. Exploiting vulnerabilities within infotainment systems, navigation history, and voice recognition features allows attackers to extract sensitive personal information. Data poisoning attacks further exacerbate security risks by manipulating AI training datasets, potentially leading to misinformation within vehicle systems, such as inaccurate route suggestions or unsafe driving decisions \cite{golda2024privacy,chamola2021information}. Techniques such as Deepfake-based identity attacks introduce another layer of vulnerability, enabling impersonation of drivers or passengers to facilitate unauthorized vehicle access or fraudulent transactions tied to biometric authentication \cite{karabulut2023privacy}. Similarly, Adversarial machine learning further amplifies security concerns by allowing attackers to exploit weaknesses in AI perception models, altering how autonomous vehicles interpret their surroundings. Manipulating road sign recognition could lead to misreading speed limits or ignoring stop signs, endangering passengers. More critically, adversarial techniques can be leveraged to extract private user data from AI models trained on driving behavior, exposing sensitive information such as frequently visited locations, daily commutes, and lifestyle preferences.} \par

The fallout from vehicle hacking also extends to automakers. A single incident can result in mass recalls, expensive software patches, and irreparable damage to the brand’s reputation. Automakers may also face legal liabilities if it is proven that they failed to implement sufficient cybersecurity protections, potentially exposing consumers to substantial risks. In short, while connected vehicles provide convenience and advanced features, they also present significant security vulnerabilities. Unauthorized access to sensitive data not only jeopardizes privacy but also poses serious safety risks, as hackers can manipulate vehicle systems or use the data for malicious purposes. Without stronger cybersecurity measures, consumers remain vulnerable to physical, financial, and emotional harm, highlighting the need for more robust protections in the rapidly evolving landscape of connected vehicles.

\subsection{Revealing Personal Travel Patterns Without Permission}
GPS systems in modern vehicles constantly track live location data to provide navigation services and identify nearby points of interest. Automakers, service providers, and navigation platforms typically access this data to enhance the driver’s experience. However, third-party companies or data aggregators may also gain access to this sensitive information indirectly, often through vague privacy policies or unauthorized partnerships. The data collected includes GPS history, preferred routes, frequently visited locations, and real-time driving behavior. When shared without the driver's consent, this data can expose personal information, revealing patterns in daily routines such as commutes, visits to medical centers, places of worship, or other personal stops.

The misuse of this information for commercial gain is a major concern \cite{salhieh2021integrating}. Marketing agencies, for example, could exploit travel data to push hyper-localized advertisements, such as promotions from nearby competitors. While some may see this as convenient, it can feel intrusive, creating the impression of constant surveillance. The thought of private routines being monitored and commodified without permission raises significant privacy concerns for consumers, leading to discomfort and distrust. More alarmingly, access to travel patterns can pose serious security risks \cite{liaw2002time}. If criminals or stalkers obtain this data, they could track an individual’s movements in real time, facilitating harassment, home invasions, or worse. For high-profile individuals or executives, this data could enable kidnapping or corporate espionage, as attackers use travel history to plan their actions \cite{idachaba2011design, johnson2006vehicle,morgan2016reducing}. There are also broader societal implications, particularly when government or law enforcement agencies gain access to this data without the driver's knowledge or proper legal oversight. This opens the door to potential abuses of power, where individuals may be monitored based on their travel behaviors without consent. This kind of surveillance undermines civil liberties, as people lose control over who tracks their location. Discrimination is another potential risk. For instance, insurance companies could use travel data to assess risk, raising premiums for those who frequently visit high-risk areas. Employers or business partners could also misuse travel history to make judgments about a person’s lifestyle or commuting habits, impacting decisions related to hiring, promotions, or collaborations.

The lack of transparency regarding how travel data is shared further complicates the issue. Many drivers unknowingly agree to terms that allow companies to collect and distribute their data without fully understanding the extent of this sharing. Once shared, the data may be sold to third parties, increasing the risk of exposure and misuse. In short, the unauthorized sharing of travel patterns from connected vehicles poses significant threats to consumer privacy, safety, and personal freedom. While the data collected offers convenience, it can be easily exploited for commercial, criminal, or governmental purposes, leading to potential harm. Stronger protections and more transparent data-sharing practices are essential to safeguard consumers and maintain trust in connected vehicle technologies.

\subsection{Financial Manipulation by Service Providers}
In today's present smart vehicles, advanced payment services, including digital wallets and built-in payment systems such as Android Auto Pay and Apple Pay, facilitate transactions for tolls, parking fees, and various services \cite{bari2022service}. While these technologies offer convenience, they also generate extensive financial data about users, which can be leveraged by third-party companies and financial service providers for various purposes, including marketing and service enhancements. However, when this data is shared, it can be exploited in several ways, leading to financial manipulation and unfair practices.

One significant concern is the potential for predictive maintenance alerts to be used manipulatively. Service providers may use real-time vehicle health data to send notifications about necessary repairs, such as oil changes or brake replacements \cite{thaduri2016context}. While this can be useful, some providers might exaggerate the urgency of repairs or suggest unnecessary services to increase their revenue. For example, a service provider could alert a driver that their brake pads are near failure, even if they still have several months of safe use remaining, pressuring the driver into spending money prematurely \cite{thaduri2016context}. Financial data linked to connected vehicles can also be exploited. Modern cars with in-car payment systems allow drivers to pay for fuel, tolls, parking, and other services directly from their vehicle \cite{gunjal2023survey}. Providers with access to this payment history might use it to inflate prices for routine services, targeting drivers who frequently use premium services with higher-priced options. For instance, if a provider notices a driver regularly purchasing high-end products, they might offer more expensive alternatives, assuming the driver is willing to pay more.

Additionally, driving behavior data can be used to create tiered pricing models for services. Service providers may segment drivers based on risk profiles and usage patterns, charging higher fees for those categorized as high-risk due to aggressive driving or frequent trips in harsh conditions. This dynamic pricing model can result in inflated service costs for minor or isolated incidents, unfairly penalizing drivers based on data analytics rather than actual driving behavior. The misuse of vehicle data can extend to influencing financial profiles and creditworthiness. For example, frequent late payments on subscription services or maintenance plans could negatively impact a driver’s credit score \cite{brockett2007biological}. Financial institutions and credit agencies might use this data to make decisions about loans and insurance, potentially leading to higher interest rates or less favorable terms \cite{brockett2007biological}. Drivers may be unaware that their vehicle-related transactions are influencing their financial profile, leaving them vulnerable to adverse financial outcomes. Another manipulation tactic involves bundling services based on driving habits and vehicle data. Service providers might offer bundled maintenance plans that appear cost-effective but are priced significantly higher than purchasing individual services. For example, a driver who commutes long distances might be offered a maintenance package that includes tire rotations and engine checks at a premium price, despite the actual cost being much lower if bought separately. Service providers may also use proprietary diagnostic data to limit access to independent repair shops. By forming exclusive agreements with automakers, they can restrict vehicle owners from seeking repairs at more affordable independent shops. This "right to repair" restriction forces drivers to use authorized service centers, often resulting in higher repair costs and less competitive pricing.

The lack of transparency in data sharing further exacerbates these issues. Consumers often agree to data-sharing terms without fully understanding the extent of the data collected and how it will be used. These opaque agreements make it challenging for drivers to protect themselves from financial manipulation, as they are left in the dark about the use of their data. In short, while connected vehicle data provides numerous conveniences, its misuse by service providers can lead to various forms of financial manipulation. From exaggerated repair alerts and inflated service costs to influencing financial profiles and limiting repair options, the exploitation of this data undermines consumer autonomy and trust. Enhanced transparency and stronger consumer protections are essential to safeguard drivers from these financial risks and ensure that connected vehicle technologies benefit users rather than exploit them.

\subsection{Increased Risk of Stalking and Harassment}
In intelligent transportation systems, vehicle and user data integration is crucial to enhancing overall system performance and user satisfaction. By combining vehicle sensor data with user preferences and filters, advanced operations such as personalized trip planning, location-based services, tailored in-vehicle services, and customized entertainment options are enabled. This integration creates a more efficient, personalized, and convenient transportation experience, aligning with individual needs in real time. However, if such data is accessed or shared without the explicit consent of the vehicle owner, it significantly heightens the risk of stalking and harassment. This unauthorized access can occur through insecure data-sharing mechanisms, data breaches, or third-party exploitation of vulnerabilities in the vehicle’s communication systems.

When connected vehicle data is compromised, it allows for detailed tracking of an individual’s movements and routines. For example, GPS data reveals frequent routes, destinations, and daily schedules, enabling malicious actors to map out a person's routine, including sensitive locations such as home and workplace. This level of tracking is particularly dangerous as it empowers stalkers or harassers to monitor the victim’s movements in real-time, granting them precise knowledge of their whereabouts at any given moment \cite{connealy2019risk}. Stalkers could discreetly use this information to intercept or follow individuals along their regular routes, potentially leading to physical confrontations, robbery, or kidnapping \cite{idachaba2011design}. Such predictability makes it easier for perpetrators to plan their actions, thereby increasing the risk of persistent harassment. Beyond physical threats, the misuse of this data facilitates psychological abuse. Stalkers may send threatening messages or unwanted communications that reference specific locations, instilling fear and manipulating victims by exploiting their intimate knowledge of the victim's movements and habits. This pervasive sense of vulnerability can lead to severe emotional distress \cite{stevens2021cyber}. Moreover, unauthorized access to vehicle data can result in broader forms of exploitation. For example, sensitive travel histories or location preferences may be leveraged by criminals for blackmail or coercion, threatening victims with exposure of personal information or further harassment. The opaque nature of how vehicle data is shared and accessed exacerbates these risks. Many vehicle owners remain unaware of how their data is collected or distributed, preventing them from proactively safeguarding their privacy. Insufficient security measures, such as storing data on insecure servers or using unprotected transmission channels, increase the likelihood of unauthorized access by cybercriminals.

Addressing the risks of stalking and harassment requires stringent data protection protocols, including encryption during data transmission and storage, and regular security audits to identify vulnerabilities. Transparent privacy policies must also inform vehicle owners about data usage and sharing, empowering them with control over their personal information. Additionally, legal and regulatory frameworks need to be updated to impose strict data protection standards and to provide recourse for consumers whose data is misused \cite{georgiadou2019location}. Strengthening legal protections against stalking and harassment, with particular emphasis on cases involving vehicle data, is essential. In short, unauthorized access to connected vehicle data greatly increases the risk of stalking and harassment, enabling malicious actors to exploit detailed tracking information for physical, psychological, and emotional harm. Implementing robust privacy measures, ensuring data transparency, and strengthening legal safeguards are crucial steps in protecting individuals from these dangers and preserving trust in connected vehicle technologies.

\subsection{Exploitation of Biometric Data Without Consent}
Data from connected vehicles, including biometric data such as facial recognition, fingerprint scans, and voice patterns, is increasingly used to enhance vehicle security, convenience, and personalization \cite{roeschlin2018bionyms}. Automakers are integrating technologies like facial recognition, voice, and iris detection to enable services such as fatigue detection, sleep alerts, and thief detection when unknown individuals enter the vehicle \cite{roeschlin2018bionyms}. However, unauthorized access or indirect sharing of sensitive biometric data may occur due to data breaches, poorly protected systems, or vague consent agreements. Biometric data typically includes unique identifiers such as fingerprints, iris scans, and voice recognition, which are stored for security purposes or personalized settings \cite{singh2010voice}. The exploitation of biometric data without consent significantly threatens consumer privacy and security. Biometric data is particularly sensitive because, unlike passwords, it cannot be easily changed. Once compromised, it remains a permanent identifier, exposing individuals to a lifetime of potential misuse. 

If unauthorized entities gain access to biometric data, it can be used to impersonate individuals, commit fraud, or bypass security systems reliant on biometric authentication \cite{ karabulut2023privacy}. For instance, if fingerprint data is accessed, hackers could unlock the vehicle or access other services linked to the fingerprint, thereby compromising not only the vehicle but also personal devices or financial accounts. The misuse of biometric data for surveillance and tracking purposes is another growing concern. Sharing biometric data with third-party companies may allow for monitoring in unanticipated ways. In some cases, biometric data might be sold to law enforcement agencies or private security firms, raising ethical concerns over civil liberties and privacy erosion. Furthermore, biometric data misuse extends to marketing and profiling. Companies may use this data to build highly detailed consumer profiles based on physical traits or emotional responses detected through biometric scans. Eye-tracking data, for example, could reveal where a driver focuses their attention while viewing advertisements, and this information could be sold to advertisers for personalized ad targeting. Voice recognition systems could detect emotional states, allowing for tailored marketing based on mood or stress levels. While convenient, these applications infringe upon consumer privacy, as users have little control over how their biometric data is used or monetized. The long-term impact of biometric data breaches is severe. Once compromised, biometric markers cannot be changed, leaving individuals vulnerable to identity theft, fraud, and exploitation for life. Biometric data remains valuable to malicious actors seeking unauthorized access or identity-related crimes. The lack of robust legal frameworks governing the collection and use of biometric data exacerbates this issue. Although some regions have implemented protections, many lack comprehensive regulations on how this data is stored, used, and shared. Consumers often face vague consent clauses in user agreements, making it difficult to understand how their data is utilized. This lack of transparency creates an environment ripe for exploitation.

To mitigate these risks, stronger regulations addressing biometric data collection and use in connected vehicles are essential. Consumers must be clearly informed about how their biometric data is collected, with transparent consent mechanisms in place. Companies should adopt robust security measures, such as encryption and secure storage protocols, to protect biometric data from unauthorized access. Legal frameworks should grant consumers the right to access, control, and delete their biometric data to ensure they maintain control over their personal information. In short, unauthorized exploitation of biometric data threatens consumer privacy, security, and autonomy, with the potential for identity theft, fraud, surveillance, and discrimination. To address these issues, stronger legal protections, transparent data usage practices, and enhanced security measures are necessary to safeguard biometric data and ensure responsible use. Without these protections, consumers remain vulnerable to the extensive consequences of biometric data misuse.

\begin{table*}[!p]
\centering
\small
\caption{\textcolor{black}{Overview of Global Data Privacy Laws and Their Relation to CAVs}}   
\begin{tabular}{|c|m{1cm}|m{2.5cm}|m{2cm}|m{5.6cm}|m{3.6cm}|}
\hline
\rowcolor[rgb]{0.949,0.949,0.949} \textcolor{black}{\textbf{No.}} \vphantom{\rule{0pt}{3.0ex}} & \begin{center} \textcolor{black}{\textbf{Region}} \end{center} & \begin{center} \textcolor{black}{\textbf{Law/Framework}} \end{center}  & \begin{center} \textcolor{black}{\textbf{Focus Area}} \end{center} & \begin{center} \textcolor{black}{\textbf{Key Features}} \end{center} & \begin{center} \textcolor{black}{\textbf{Relation to CAVs}} \end{center} \\ 
\hline
\rowcolor[rgb]{0.902,0.925,0.949} \textcolor{black}{\textbf{1.}} & 
\textcolor{black}{European Union (EU)}  & \textcolor{black}{General Data Protection Regulation (GDPR) \cite{gdpr2016general}} &  \textcolor{black}{Data protection rights, consent, and transparency} &\textcolor{black}{{\begin{enumerate}
\item[-] Comprehensive data privacy regulation. 
\item[-] Requires explicit consent and transparency. 
\item[-] Data breach reporting within 72 hours.
\end{enumerate}}} 
& \textcolor{black}{Strong relevance due to high consumer protections. Mandates transparency in data collected by vehicles, e.g., GPS and biometrics.} \\ 
\hline
\rowcolor[rgb]{0.902,0.925,0.949}  \textcolor{black}{\textbf{2.}} & \textcolor{black}{European Union (EU)} & \textcolor{black}{ePrivacy Directive (Cookie Law) \cite{bond2012eu}} & \textcolor{black}{Privacy in electronic communications} & 
\textcolor{black}{{\begin{enumerate}
\item[-] Regulates cookies and electronic marketing.  
\item[-] Works alongside GDPR.\end{enumerate}}
}
& \textcolor{black}{Focuses on communication systems within connected cars, e.g., infotainment systems and mobile app integration.} \\ 
\hline
\textcolor{black}{\textbf{3.}} & \textcolor{black}{USA (California)} & \textcolor{black}{California Consumer Privacy Act (CCPA) \cite{harding2019understanding}} & \textcolor{black}{Consumer rights over personal information} & {
\textcolor{black}{\begin{enumerate}\item[-] Rights to access, delete, and opt-out of the sale of personal data. 
\item[-] Extended by CPRA to include more stringent protections.\end{enumerate}}
} & \textcolor{black}{Empowers consumers to control personal data collected by cars, such as location tracking, driving behavior, and in-car sensors.} \\ 
\hline
\rowcolor[rgb]{0.902,0.925,0.949} \textcolor{black}{\textbf{4.}} & \textcolor{black}{Canada} & \textcolor{black}{Personal Information Protection and Electronic Documents Act (PIPEDA) \cite{jaar2008canadian}} & \textcolor{black}{Private-sector data protection} & {
\textcolor{black}{\begin{enumerate}\item[-] Governs the collection, use, and disclosure of personal data. 
\item[-] Requires consent for data use.\end{enumerate}} 
} & \textcolor{black}{Regulates automakers and third-party apps collecting driver data, ensuring secure and consent-based handling of connected car data.} \\ 
\hline
\rowcolor[rgb]{0.902,0.925,0.949}
\textcolor{black}{\textbf{5.}} & \textcolor{black}{Canada} & \textcolor{black}{Consumer Privacy Protection Act (CPPA) \cite{schwanen2023getting}} & \textcolor{black}{Strengthening privacy laws} & 
\textcolor{black}{%
\begin{enumerate}
\item[-] Aims to modernize and strengthen privacy laws. 
\item[-] Proposes higher penalties for non-compliance.
\end{enumerate}}   
& \textcolor{black}{Could offer improved safeguards for vehicle data, including clarity on who owns the data collected by autonomous vehicles.} \\ 
\hline
\textcolor{black}{\textbf{6.}} & \textcolor{black}{Singapore} & \textcolor{black}{Personal Data Protection Act (PDPA) \cite{chik2013singapore}} & \textcolor{black}{Collection, use, and disclosure of personal data} &
{
\textcolor{black}{\begin{enumerate}\item[-] Requires clear purpose for data collection. 
\item[-] Data breach notification mandatory. - Offers right to access and correct data.\end{enumerate}}}
& \textcolor{black}{Ensures consent for collection of location, biometric, and behavior data in connected cars sold or operated in Singapore.} \\ 
\hline
\rowcolor[rgb]{0.902,0.925,0.949}
\textcolor{black}{ \textbf{7.}} & \textcolor{black}{China} & \textcolor{black}{Data Security Law (DSL) \cite{chen2021understanding}} & \textcolor{black}{Data sovereignty and security} & {
\textcolor{black}{\begin{enumerate}\item[-] Focus on national security. 
\item[-] Data classification and protection levels. 
\item[-] Restricts cross-border data transfers.\end{enumerate}} 
}
& \textcolor{black}{Aims to ensure vehicle data, such as mapping and behavior analytics, remains secure and within China's borders.} \\ 
\hline
\rowcolor[rgb]{0.902,0.925,0.949} \textcolor{black}{\textbf{8.}} & \textcolor{black}{China} & \textcolor{black}{Personal Information Protection Law (PIPL) \cite{calzada2022citizens,zhou2024understanding}} & \textcolor{black}{Data privacy, inspired by GDPR} & {
\textcolor{black}{\begin{enumerate}\item[-] Rights for access, correction, and deletion. \item[-] Requires explicit consent for data processing.
\item[-] Strict rules for cross-border data transfers.\end{enumerate}}
}
& \textcolor{black}{Offers strict protections for data collected by connected cars, addressing concerns like invasive surveillance or third-party sharing.} \\ 
\hline
\textcolor{black}{\textbf{9.}} & \textcolor{black}{Australia} & \textcolor{black}{Privacy Act 1988 \cite{yuvaraj2018me,taylor2020personal}} & \textcolor{black}{Australian Privacy Principles (APPs)} & {
\textcolor{black}{\begin{enumerate}\item[-] 13 APPs govern data collection, storage, and usage.
\item[-] Mandatory data breach notification.\end{enumerate}}
}
& \textcolor{black}{Requires automakers to handle driver data responsibly and notify consumers in case of data breaches.} \\ 
\hline
\rowcolor[rgb]{0.902,0.925,0.949} \textcolor{black}{\textbf{10.}} & \textcolor{black}{Qatar} & \textcolor{black}{Personal Data Privacy Protection Law \cite{qatarprivacy2016}} & \textcolor{black}{Comprehensive data protection} & {
\textcolor{black}{\begin{enumerate}\item[-] Protects individual rights concerning privacy and personal data. 
\item[-] Restricts direct marketing and protection for processing children's data (e.g., parental consent).\end{enumerate}}
}
& \textcolor{black}{Applicable to automakers and service providers operating in Qatar.}  \\ 
\hline
\end{tabular}
\begin{center}
\label{tab:data_privacy_laws_cars}
\end{center}
\end{table*}

\subsection{Linking Vehicle Data with Other Personal Devices}
Modern vehicles increasingly rely on the integration of users' personal devices to enable various features and services, such as hands-free calling, controlling navigation apps, playing media from personal playlists, sending and receiving messages, processing payments, and activating voice-activated controls via virtual assistants. To facilitate these services, users' smartphones, tablets, or other devices connect to the vehicle using Bluetooth, Wi-Fi, or USB. This connection enables the creation of a comprehensive and detailed user profile by linking vehicle data with data from personal devices \cite{reininger2015first}. As a result, privacy concerns arise when automakers, app developers, and third-party companies gain access to and cross-reference personal information across platforms. For instance, a driver's smartphone might share GPS location data with the vehicle, which could then sync with other personal devices like a smartwatch monitoring physical activity. This convergence of data generates a detailed profile of the user's habits, locations, health information, and communication patterns, which may be used for purposes beyond the driver's original consent.

Furthermore, linking vehicle data with personal devices amplifies security risks, as each connected device becomes a potential vulnerability. If one device is compromised, it could expose the entire network, including the vehicle. A hacker who gains access to a driver's smartphone might be able to manipulate the vehicle's settings, track its location, or even disable critical systems like the alarm \cite{palm2021ethical}. This level of control could lead to more severe outcomes, such as unauthorized access to personal information, identity theft, or physical harm if vehicle systems are tampered with. Additionally, companies may misuse this interconnected data to create highly detailed consumer profiles, exploiting them for commercial gain. For example, an automaker could collect a driver's health metrics from their smartwatch and combine this data with driving behavior and location information. Insurers might adjust premiums based on perceived health risks or driving patterns, while advertisers could target the driver with specific health or lifestyle products. Such aggregation of personal data across multiple platforms creates an environment where individuals have limited control over how their data is used, while companies profit from the data without offering tangible benefits to the user.

In short, linking vehicle data with personal devices opens up numerous pathways for data exploitation and security breaches. The aggregation of personal information across various platforms enables companies to construct detailed consumer profiles that can be misused for commercial purposes or result in privacy violations. To safeguard consumers, robust data protection measures, transparent consent practices, and updated regulatory frameworks must be implemented to account for the complexities of modern, interconnected systems. Without these protections, the integration of vehicle data with personal devices will continue to erode consumer trust and expose individuals to significant risks.


\section{Regulatory Framework and Standards}
\label{Section: Regulatory Framework and Standards}

\subsection{\textcolor{black}{Regional Variations in Data-Sharing Mechanisms}}
\textcolor{black}{Data-sharing practices in CAVs exhibit significant regional variations, influenced by regulatory frameworks, infrastructure readiness, and technological maturity. Distinct approaches have emerged, reflecting the disparities in governance structures and the extent of digital integration within transportation ecosystems.}
\begin{enumerate}
    \item \textcolor{black}{North America (U.S. \& Canada):
    In the United States, a fragmented approach to data privacy governs automotive data, with regulations like the California Consumer Privacy Act (CCPA) shaping data governance practices. The absence of a unified federal policy introduces challenges for automakers, resulting in inconsistent data-sharing protocols across states. While private sector-driven innovation, exemplified by companies such as Tesla and Waymo, has spurred the adoption of CAVs, concerns regarding user data security persist.}

\item \textcolor{black}{Europe (EU \& UK):
    The General Data Protection Regulation (GDPR) enforces strict data privacy policies, mandating user consent and ensuring transparency in data collection processes. Initiatives such as Gaia-X seek to establish a standardized and secure data-sharing infrastructure across European countries. Automakers are required to adopt privacy-by-design principles, which limit excessive data collection and bolster cybersecurity measures to safeguard user information.}

\item \textcolor{black}{Asia-Pacific (China, Japan, South Korea):
China enforces strict state-controlled data governance, mandating that vehicle-generated data be stored within national borders. Japan and South Korea utilize advanced 5G and AI-driven smart mobility initiatives, supported by government-led efforts to ensure secure vehicle-to-infrastructure (V2I) communication. In certain regions, blockchain-based solutions are emerging to strengthen data security and authentication within automotive ecosystems \cite{alladi2022comprehensive}.}
\end{enumerate}

\subsection{Regulatory Framework and Standards Over Worldwide}
Currently, to address concerns related to user data protection and privacy, numerous countries have introduced specific laws and regulatory frameworks. In the European Union (EU), the General Data Protection Regulation (GDPR) \cite{gdpr2016general} serves as a comprehensive and stringent standard for data privacy, ensuring individuals have greater control over their personal data. Similarly, in the United States, the California Consumer Privacy Act (CCPA) \cite{harding2019understanding} establishes key protections for residents of California, granting them rights to access, delete, and control their personal information collected by businesses. Following these examples, many other nations have enacted their own laws tailored to their jurisdictions, reflecting the growing global emphasis on safeguarding user data and ensuring accountability for entities handling personal information. Table \ref{tab:data_privacy_laws_cars} provides an overview of various data privacy laws and frameworks introduced worldwide. 

However, in reality, no country has specifically tailored its legal frameworks to address data sharing and user or passenger privacy in the automobile sector, particularly for CAVs. While general data protection laws such as GDPR and CCPA provide overarching guidelines for data privacy, they fail to address the unique challenges posed by the automotive industry, where vehicles constantly generate and transmit vast amounts of sensitive data. This includes location information, driving patterns, in-vehicle communications, and biometric data of passengers, all of which are susceptible to misuse or unauthorized access. The lack of specific regulations leaves significant gaps in safeguarding user privacy in an era where CAVs technologies are becoming integral to modern transportation systems. 

The absence of specific laws, frameworks, and regulatory oversight for data sharing in the automotive sector has created a significant privacy vacuum. Vehicle manufacturers and third-party companies are exploiting this gap by collecting users' data without their explicit consent and often without informing them. This unregulated data collection frequently exceeds the bounds of what is necessary for the vehicle's core functionalities, venturing into excessive and intrusive data harvesting practices. Moreover, these entities retain collected data far beyond the necessary timeframes, increasing the risk of misuse, unauthorized access, or breaches. This lack of accountability undermines user trust, compromises privacy, and highlights the urgent need for targeted regulations to ensure ethical and lawful handling of data in the CAVs ecosystems.

The lack of transparency in data sharing compounds these issues, leaving consumers in a vulnerable position. Most users unknowingly consent to vague or convoluted data-sharing terms, often buried within lengthy agreements, without comprehending the breadth of data being collected or its intended use. This opacity not only prevents users from making informed decisions about their privacy but also enables manufacturers and third-party companies to exploit this lack of awareness for unrestricted data collection. Consequently, sensitive information, such as real-time location, behavioral patterns, and even biometric data, can be harvested and utilized in ways that consumers neither anticipate nor approve, further undermining trust and exposing them to potential risks.

To address these concerns, there is a pressing need for regulatory and standardization bodies at both the global and national levels. These entities must establish stringent guidelines and rules to ensure that vehicle manufacturers and third-party companies operate transparently and ethically in their data practices. Such regulations should mandate that no user data is collected without explicit, informed consent and that the scope of data collection is strictly limited to what is necessary for the vehicle's functionality. Additionally, these frameworks should enforce stringent penalties for violations, require data minimization, and establish clear policies for data retention and deletion. By implementing robust oversight mechanisms, these regulatory bodies can safeguard consumer privacy, foster trust, and ensure that the rapid evolution of CAVs aligns with ethical and legal standards.
 

\section{Conclusion}
\label{Section: Conclusion}
The advent of CAVs has redefined modern mobility, delivering unparalleled advantages such as enhanced safety, real-time navigation, and personalized user experiences through sophisticated data-sharing networks like V2V, V2I, and V2X. While these innovations promise significant consumer benefits, they also usher in a new era of privacy concerns, transforming what could be a consumer advantage into a potential privacy nightmare. The extensive collection and sharing of user data—spanning location, behavior, and personal identifiers—expose users to the risk of unauthorized access and exploitation by third parties, often without explicit consent. Current data protection regulations, while valuable, fail to address the unique complexities of CAV ecosystems, leaving critical gaps in privacy safeguards. To protect consumers and foster trust, it is imperative to develop robust, domain-specific regulations that prioritize user consent, transparency, and accountability. These policies should encompass strict data usage protocols, clear retention and deletion mechanisms, and enforceable penalties for violations. By aligning technological innovation with ethical responsibility, the automotive industry can ensure that the benefits of CAVs do not come at the cost of consumer privacy. Striking this balance will not only secure sensitive user data but also pave the way for a future where consumers can fully embrace these transformative technologies without compromising their trust or privacy.

\bibliographystyle{IEEEtranN}
{\footnotesize
\bibliography{references.bib}}

@inproceedings{kamalkumar2022lidar,
  title={LIDAR Based Self Driving Car},
  author={Kamalkumar, V and Sonali, S and Sindhuja, V and Sriharish, A},
  booktitle={Proceedings of the International Conference on Intelligent Technologies in Security and Privacy for Wireless Communication, ITSPWC 2022, 14-15 May 2022, Karur, Tamilnadu, India},
  year={2022}
}

@article{abbasi2022lidar,
  title={Lidar point cloud compression, processing and learning for autonomous driving},
  author={Abbasi, Rashid and Bashir, Ali Kashif and Alyamani, Hasan J and Amin, Farhan and Doh, Jaehyeok and Chen, Jianwen},
  journal={IEEE Transactions on Intelligent Transportation Systems},
  volume={24},
  number={1}, 
  pages={962--979},
  year={2022},
  publisher={IEEE}
}

@inproceedings{li2017smart,
  title={Smart home energy management with vehicle-to-home technology},
  author={Li, Chenxi and Luo, Fengji and Chen, Yingying and Xu, Zhao and An, Yinan and Li, Xiao},
  booktitle={2017 13th IEEE international conference on control \& automation (ICCA)},
  pages={136--142},
  year={2017},
  organization={IEEE}
}

@article{hussain2022drivable,
  title={Drivable region estimation for self-driving vehicles using radar},
  author={Hussain, Muhammad Ishfaq and Azam, Shoaib and Rafique, Muhammad Aasim and Sheri, Ahmad Muqeem and Jeon, Moongu},
  journal={IEEE Transactions on Vehicular Technology},
  volume={71},
  number={6},
  pages={5971--5982},
  year={2022},
  publisher={IEEE}
}

@inproceedings{hussain2020multiple,
  title={Multiple objects tracking using radar for autonomous driving},
  author={Hussain, Muhamamd Ishfaq and Azam, Shoaib and Munir, Farzeen and Khan, Zafran and Jeon, Moongu},
  booktitle={2020 IEEE International IOT, Electronics and Mechatronics Conference (IEMTRONICS)},
  pages={1--4},
  year={2020},
  organization={IEEE}
}

@article{liu2021robust,
  title={Robust target recognition and tracking of self-driving cars with radar and camera information fusion under severe weather conditions},
  author={Liu, Ze and Cai, Yingfeng and Wang, Hai and Chen, Long and Gao, Hongbo and Jia, Yunyi and Li, Yicheng},
  journal={IEEE Transactions on Intelligent Transportation Systems},
  volume={23},
  number={7},
  pages={6640--6653},
  year={2021},
  publisher={IEEE}
}

@inproceedings{rashed2021bev,
  title={Bev-modnet: Monocular camera based bird's eye view moving object detection for autonomous driving},
  author={Rashed, Hazem and Essam, Mariam and Mohamed, Maha and Sallab, Ahmad Ei and Yogamani, Senthil},
  booktitle={2021 IEEE International Intelligent Transportation Systems Conference (ITSC)},
  pages={1503--1508},
  year={2021},
  organization={IEEE}
}

@inproceedings{zhao2020dynamic,
  title={Dynamic object tracking for self-driving cars using monocular camera and lidar},
  author={Zhao, Lin and Wang, Meiling and Su, Sheng and Liu, Tong and Yang, Yi},
  booktitle={2020 IEEE/RSJ International Conference on Intelligent Robots and Systems (IROS)},
  pages={10865--10872},
  year={2020},
  organization={IEEE}
}

@article{zaarane2020distance,
  title={Distance measurement system for autonomous vehicles using stereo camera},
  author={Zaarane, Abdelmoghit and Slimani, Ibtissam and Al Okaishi, Wahban and Atouf, Issam and Hamdoun, Abdellatif},
  journal={Array},
  volume={5},
  pages={100016},
  year={2020},
  publisher={Elsevier}
}

@article{bhadoriya2022vehicle,
  title={Vehicle detection and tracking using thermal cameras in adverse visibility conditions},
  author={Bhadoriya, Abhay Singh and Vegamoor, Vamsi and Rathinam, Sivakumar},
  journal={Sensors},
  volume={22},
  number={12},
  pages={4567},
  year={2022},
  publisher={MDPI}
}

@inproceedings{shin2023deep,
  title={Deep depth estimation from thermal image},
  author={Shin, Ukcheol and Park, Jinsun and Kweon, In So},
  booktitle={Proceedings of the IEEE/CVF Conference on Computer Vision and Pattern Recognition},
  pages={1043--1053},
  year={2023}
}

@article{kumar2023surround,
  title={Surround-view fisheye camera perception for automated driving: Overview, survey \& challenges},
  author={Kumar, Varun Ravi and Eising, Ciar{\'a}n and Witt, Christian and Yogamani, Senthil Kumar},
  journal={IEEE Transactions on Intelligent Transportation Systems},
  volume={24},
  number={4},
  pages={3638--3659},
  year={2023},
  publisher={IEEE}
}

@article{xu2018analyzing,
  title={Analyzing and enhancing the security of ultrasonic sensors for autonomous vehicles},
  author={Xu, Wenyuan and Yan, Chen and Jia, Weibin and Ji, Xiaoyu and Liu, Jianhao},
  journal={IEEE Internet of Things Journal},
  volume={5},
  number={6},
  pages={5015--5029},
  year={2018},
  publisher={IEEE}
}

@article{wong2020mapping,
  title={Mapping for autonomous driving: Opportunities and challenges},
  author={Wong, Kelvin and Gu, Yanlei and Kamijo, Shunsuke},
  journal={IEEE Intelligent Transportation Systems Magazine},
  volume={13},
  number={1},
  pages={91--106},
  year={2020},
  publisher={IEEE}
}

@article{mohamed2019survey,
  title={A survey on odometry for autonomous navigation systems},
  author={Mohamed, Sherif AS and Haghbayan, Mohammad-Hashem and Westerlund, Tomi and Heikkonen, Jukka and Tenhunen, Hannu and Plosila, Juha},
  journal={IEEE access},
  volume={7},
  pages={97466--97486},
  year={2019},
  publisher={IEEE}
}

@article{brunker2018odometry,
  title={Odometry 2.0: A slip-adaptive EIF-based four-wheel-odometry model for parking},
  author={Brunker, Alexander and Wohlgemuth, Thomas and Frey, Michael and Gauterin, Frank},
  journal={IEEE Transactions on Intelligent Vehicles},
  volume={4},
  number={1},
  pages={114--126},
  year={2018},
  publisher={IEEE}
}

@article{thakur2018infrared,
  title={Infrared sensors for autonomous vehicles},
  author={Thakur, Rajeev},
  journal={Recent Development in Optoelectronic Devices},
  volume={29},
  year={2018},
  publisher={BoD- Books on Demand: London}
}

@incollection{dong2024sensors,
  title={Sensors for autonomous vehicles},
  author={Dong, Weiqiang},
  booktitle={Handbook of Power Electronics in Autonomous and Electric Vehicles},
  pages={29--43},
  year={2024},
  publisher={Elsevier}
}

@article{nazemipour2020mems,
  title={MEMS gyro bias estimation in accelerated motions using sensor fusion of camera and angular-rate gyroscope},
  author={Nazemipour, Ali and Manzuri, Mohammad Taghi and Kamran, Danial and Karimian, Mahdi},
  journal={IEEE Transactions on Vehicular Technology},
  volume={69},
  number={4},
  pages={3841--3851},
  year={2020},
  publisher={IEEE}
}

@article{zylius2017investigation,
  title={Investigation of route-independent aggressive and safe driving features obtained from accelerometer signals},
  author={Zylius, Gediminas},
  journal={IEEE Intelligent Transportation Systems Magazine},
  volume={9},
  number={2},
  pages={103--113},
  year={2017},
  publisher={IEEE}
}

@inproceedings{braun2015capseat,
  title={CapSeat: capacitive proximity sensing for automotive activity recognition},
  author={Braun, Andreas and Frank, Sebastian and Majewski, Martin and Wang, Xiaofeng},
  booktitle={Proceedings of the 7th International Conference on Automotive User Interfaces and Interactive Vehicular Applications},
  pages={225--232},
  year={2015}
}

@article{nguyen2023risk,
  title={Risk-informed decision-making and control strategies for autonomous vehicles in emergency situations},
  author={Nguyen, Hung Duy and Choi, Mooryong and Han, Kyoungseok},
  journal={Accident Analysis \& Prevention},
  volume={193},
  pages={107305},
  year={2023},
  publisher={Elsevier}
}

@inproceedings{liu2022towards,
  title={Towards A Real-Time Emergency Response Model For Connected And Autonomous Vehicles.},
  author={Liu, Yen-Hung and de Paula Albuquerque, Ot{\'a}vio and Hung, Patrick CK and Gabbar, Hossam A and Fantinato, Marcelo and Iqbal, Farkhund},
  booktitle={CIKM Workshops},
  year={2022}
}

@article{chougule2023comprehensive,
  title={A comprehensive review on limitations of autonomous driving and its impact on accidents and collisions},
  author={Chougule, Amit and Chamola, Vinay and Sam, Aishwarya and Yu, Fei Richard and Sikdar, Biplab},
  journal={IEEE Open Journal of Vehicular Technology},
  year={2023},
  publisher={IEEE}
}

@article{tseng2017personalized,
  title={Personalized prediction of vehicle energy consumption based on participatory sensing},
  author={Tseng, Chien-Ming and Chau, Chi-Kin},
  journal={IEEE Transactions on Intelligent Transportation Systems},
  volume={18},
  number={11},
  pages={3103--3113},
  year={2017},
  publisher={IEEE}
}

@article{liao2024review,
  title={A Review of Personalization in Driving Behavior: Dataset, Modeling, and Validation},
  author={Liao, Xishun and Zhao, Zhouqiao and Barth, Matthew J and Abdelraouf, Amr and Gupta, Rohit and Han, Kyungtae and Ma, Jiaqi and Wu, Guoyuan},
  journal={IEEE Transactions on Intelligent Vehicles},
  year={2024},
  publisher={IEEE}
}

@article{mantouka2022deep,
  title={Deep reinforcement learning for personalized driving recommendations to mitigate aggressiveness and riskiness: Modeling and impact assessment},
  author={Mantouka, Eleni G and Vlahogianni, Eleni I},
  journal={Transportation research part C: emerging technologies},
  volume={142},
  pages={103770},
  year={2022},
  publisher={Elsevier}
}

@inproceedings{rogers1998personalized,
  title={Personalized driving route recommendations},
  author={Rogers, Seth and Langley, Pat},
  booktitle={Proceedings of the American Association of Artificial Intelligence Workshop on Recommender Systems},
  pages={96--100},
  year={1998}
}

@article{ge2019route,
  title={Route recommendations for intelligent transportation services},
  author={Ge, Yong and Li, Huayu and Tuzhilin, Alexander},
  journal={IEEE Transactions on Knowledge and Data Engineering},
  volume={33},
  number={3},
  pages={1169--1182},
  year={2019},
  publisher={IEEE}
}

@article{gilman2015personalised,
  title={Personalised assistance for fuel-efficient driving},
  author={Gilman, Ekaterina and Keskinarkaus, Anja and Tamminen, Satu and Pirttikangas, Susanna and R{\"o}ning, Juha and Riekki, Jukka},
  journal={Transportation Research Part C: Emerging Technologies},
  volume={58},
  pages={681--705},
  year={2015},
  publisher={Elsevier}
}

@article{rios2019fuel,
  title={Fuel consumption for various driving styles in conventional and hybrid electric vehicles: Integrating driving cycle predictions with fuel consumption optimization},
  author={Rios-Torres, Jackeline and Liu, Jun and Khattak, Asad},
  journal={International Journal of Sustainable Transportation},
  volume={13},
  number={2},
  pages={123--137},
  year={2019},
  publisher={Taylor \& Francis}
}

@article{solanki2017iot,
  title={An IoT based predictive connected car maintenance approach},
  author={Solanki, Vijender Kumar and Dhall, Rohit},
  year={2017},
  publisher={International Journal of Interactive Multimedia and Artificial Intelligence~…}
}

@techreport{bickelhaupt2024towards,
  title={Towards Future Vehicle Diagnostics in Software-Defined Vehicles},
  author={Bickelhaupt, Sandra and Hahn, Michael and Morozov, Andrey and Weyrich, Michael},
  year={2024},
  institution={SAE Technical Paper}
}

@article{rognvaldsson2018self,
  title={Self-monitoring for maintenance of vehicle fleets},
  author={R{\"o}gnvaldsson, Thorsteinn and Nowaczyk, S{\l}awomir and Byttner, Stefan and Prytz, Rune and Svensson, Magnus},
  journal={Data mining and knowledge discovery},
  volume={32},
  pages={344--384},
  year={2018},
  publisher={Springer}
}

@article{jeong2018integrated,
  title={An integrated self-diagnosis system for an autonomous vehicle based on an IoT gateway and deep learning},
  author={Jeong, YiNa and Son, SuRak and Jeong, EunHee and Lee, ByungKwan},
  journal={Applied Sciences},
  volume={8},
  number={7},
  pages={1164},
  year={2018},
  publisher={MDPI}
}

@book{denton2020advanced,
  title={Advanced automotive fault diagnosis: automotive technology: vehicle maintenance and repair},
  author={Denton, Tom},
  year={2020},
  publisher={Routledge}
}

@inproceedings{shen2012real,
  title={Real-time road traffic fusion and prediction with GPS and fixed-sensor data},
  author={Shen, Wei and Wynter, Laura},
  booktitle={2012 15th International Conference on Information Fusion},
  pages={1468--1475},
  year={2012},
  organization={IEEE}
}

@article{chen2020short,
  title={A short-term traffic prediction model in the vehicular cyber--physical systems},
  author={Chen, Chen and Liu, Xiaomin and Qiu, Tie and Sangaiah, Arun Kumar},
  journal={Future Generation Computer Systems},
  volume={105},
  pages={894--903},
  year={2020},
  publisher={Elsevier}
}

@article{azimjonov2021real,
  title={A real-time vehicle detection and a novel vehicle tracking systems for estimating and monitoring traffic flow on highways},
  author={Azimjonov, Jahongir and {\"O}zmen, Ahmet},
  journal={Advanced Engineering Informatics},
  volume={50},
  pages={101393},
  year={2021},
  publisher={Elsevier}
}

@article{wang2020truck,
  title={Truck traffic flow prediction based on LSTM and GRU methods with sampled GPS data},
  author={Wang, Shengyou and Zhao, Jin and Shao, Chunfu and Dong, Chunjiao and Yin, Chaoying},
  journal={Ieee Access},
  volume={8},
  pages={208158--208169},
  year={2020},
  publisher={IEEE}
}

@article{gahlan2016gps,
  title={GPS based parking system},
  author={Gahlan, Mamta and Malik, Vinita and Kaushik, Dheeraj and others},
  journal={Compusoft},
  volume={5},
  number={1},
  pages={2053--2056},
  year={2016},
  publisher={COMPUSOFT, An international journal of advanced computer technology}
}

@inproceedings{tripathi2020smart,
  title={Smart vehicle parking system using IoT},
  author={Tripathi, Vinay Raj},
  booktitle={2020 International Conference on Electrical and Electronics Engineering (ICE3)},
  pages={285--290},
  year={2020},
  organization={IEEE}
}

@article{parmar2020study,
  title={Study on demand and characteristics of parking system in urban areas: A review},
  author={Parmar, Janak and Das, Pritikana and Dave, Sanjaykumar M},
  journal={Journal of Traffic and Transportation Engineering (English Edition)},
  volume={7},
  number={1},
  pages={111--124},
  year={2020},
  publisher={Elsevier}
}

@article{chougule2023novel,
  title={A Novel Framework for Traffic Congestion Management at Intersections Using Federated Learning and Vertical partitioning},
  author={Chougule, Amit and Chamola, Vinay and Hassija, Vikas and Gupta, Pranav and Yu, F Richard},
  journal={IEEE Transactions on Consumer Electronics},
  year={2023},
  publisher={IEEE}
}

@article{chakrabarty2013analysis,
  title={Analysis of driver behaviour and crash characteristics during adverse weather conditions},
  author={Chakrabarty, Neelima and Gupta, Kamini},
  journal={Procedia-social and behavioral sciences},
  volume={104},
  pages={1048--1057},
  year={2013},
  publisher={Elsevier}
}

@article{adomah2022safety,
  title={Safety impact of connected vehicles on driver behavior in rural work zones under foggy weather conditions},
  author={Adomah, Eric and Khoda Bakhshi, Arash and Ahmed, Mohamed M},
  journal={Transportation research record},
  volume={2676},
  number={3},
  pages={88--107},
  year={2022},
  publisher={SAGE Publications Sage CA: Los Angeles, CA}
}

@techreport{ahmed2018driver,
  title={Driver performance and behavior in adverse weather conditions: An investigation using the SHRP2 naturalistic driving study data—Phase 2},
  author={Ahmed, Mohamed and Ghasemzadeh, Ali and Hammit, Britton E and Khan, Nasim and Das, Anik and Ali, Elhashemi and Young, Rhonda and Eldeeb, Hesham and others},
  year={2018},
  institution={Wyoming. Dept. of Transportation}
}

@book{daly2011automotive,
  title={Automotive air conditioning and climate control systems},
  author={Daly, Steven},
  year={2011},
  publisher={Elsevier}
}

@inproceedings{wang2018model,
  title={Model predictive climate control of connected and automated vehicles for improved energy efficiency},
  author={Wang, Hao and Kolmanovsky, Ilya and Amini, Mohammad Reza and Sun, Jing},
  booktitle={2018 Annual American Control Conference (ACC)},
  pages={828--833},
  year={2018},
  organization={IEEE}
}

@incollection{kadas2013role,
  title={The Role of Roadside Assistance in Vehicular Communication Networks: Security, Quality of Service, and Routing Issues},
  author={Kadas, George and Chatzimisios, Periklis},
  booktitle={Roadside Networks for Vehicular Communications: Architectures, Applications, and Test Fields},
  pages={1--37},
  year={2013},
  publisher={IGI Global}
}

@article{chand2020analysis,
  title={Analysis of vehicle breakdown frequency: a case study of New South Wales, Australia},
  author={Chand, Sai and Moylan, Emily and Waller, S Travis and Dixit, Vinayak},
  journal={Sustainability},
  volume={12},
  number={19},
  pages={8244},
  year={2020},
  publisher={MDPI}
}

@article{goregaonkar2013safe,
  title={Safe Driving and Accidental Monitoring Using GPS System and Three Axis Accelerometer},
  author={Goregaonkar, Roma and Bhosale, Snehal},
  journal={International Journal of Emerging Technology and Advanced Engineering},
  volume={3},
  number={11},
  year={2013},
  publisher={Citeseer}
}

@inproceedings{deserno2020transforming,
  title={Transforming smart vehicles and smart homes into private diagnostic spaces},
  author={Deserno, Thomas M},
  booktitle={Proceedings of the 2020 2nd Asia Pacific Information Technology Conference},
  pages={165--171},
  year={2020}
}

@inproceedings{yaici2022internet,
  title={Internet of Things (IoT)-based system for smart home heating and cooling control},
  author={Ya{\"\i}ci, Wahiba and Entchev, Evgueniy and Longo, Michela},
  booktitle={2022 IEEE International Conference on Environment and Electrical Engineering and 2022 IEEE Industrial and Commercial Power Systems Europe (EEEIC/I\&CPS Europe)},
  pages={1--6},
  year={2022},
  organization={IEEE}
}

@article{torad2022smart,
  title={Smart garage utilizing internet of things (IoT)},
  author={Torad, Mohamed A and Bouallegue, Belgacem and Khattab, Mahmoud M and Ahmed, Abdelmoty M},
  journal={Journal of Sensors},
  volume={2022},
  number={1},
  pages={9070683},
  year={2022},
  publisher={Wiley Online Library}
}

@inproceedings{chen2016multi,
  title={Multi-function vehicle and control method for intelligent home security environment monitoring},
  author={Chen, Yang and Deng, Yanni and Xiao, Wenchao and Hao, Fan},
  booktitle={2016 4th International Conference on Mechanical Materials and Manufacturing Engineering},
  pages={671--674},
  year={2016},
  organization={Atlantis Press}
}

@article{tian2016real,
  title={Real-time charging station recommendation system for electric-vehicle taxis},
  author={Tian, Zhiyong and Jung, Taeho and Wang, Yi and Zhang, Fan and Tu, Lai and Xu, Chengzhong and Tian, Chen and Li, Xiang-Yang},
  journal={IEEE Transactions on Intelligent Transportation Systems},
  volume={17},
  number={11},
  pages={3098--3109},
  year={2016},
  publisher={IEEE}
}

@article{lin2024pdqn,
  title={PDQN: User Preference-Based Charging Station Recommendation},
  author={Lin, Hai and Li, Xiaoyu and Cao, Yue and Labiod, Houda and Ahmad, Naveed},
  journal={IEEE Transactions on Consumer Electronics},
  year={2024},
  publisher={IEEE}
}

@inproceedings{ccano2017mood,
  title={Mood-based on-car music recommendations},
  author={{\c{C}}ano, Erion and Coppola, Riccardo and Gargiulo, Eleonora and Marengo, Marco and Morisio, Maurizio},
  booktitle={Industrial Networks and Intelligent Systems: Second International Conference, INISCOM 2016, Leicester, UK, October 31--November 1, 2016, Proceedings 2},
  pages={154--163},
  year={2017},
  organization={Springer}
}

@article{wang2021restaurant,
  title={Restaurant recommendation in vehicle context based on prediction of traffic conditions},
  author={Wang, Zehong and Liu, Jianhua and Shen, Shigen and Li, Minglu},
  journal={International Journal of Pattern Recognition and Artificial Intelligence},
  volume={35},
  number={10},
  pages={2159044},
  year={2021},
  publisher={World Scientific}
}

@article{shambour2023restaurant,
  title={Restaurant Recommendations Based on Multi-Criteria Recommendation Algorithm},
  author={Shambour, Qusai Y and Abualhaj, Mosleh M and Abu-Shareha, Ahmad Adel},
  journal={Journal of Universal Computer Science},
  volume={29},
  number={2},
  pages={179},
  year={2023},
  publisher={Pensoft Publishers}
}

@article{prytz2015predicting,
  title={Predicting the need for vehicle compressor repairs using maintenance records and logged vehicle data},
  author={Prytz, Rune and Nowaczyk, S{\l}awomir and R{\"o}gnvaldsson, Thorsteinn and Byttner, Stefan},
  journal={Engineering applications of artificial intelligence},
  volume={41},
  pages={139--150},
  year={2015},
  publisher={Elsevier}
}

@article{mills2019changing,
  title={Changing patterns of motor vehicle collision risk during winter storms: A new look at a pervasive problem},
  author={Mills, Brian and Andrey, Jean and Doberstein, Brent and Doherty, Sean and Yessis, Jennifer},
  journal={Accident Analysis \& Prevention},
  volume={127},
  pages={186--197},
  year={2019},
  publisher={Elsevier}
}

@article{malmivuo2017effects,
  title={Effects of winter Tyre type on roughness and polishing of road surfaces covered with ice and compact snow},
  author={Malmivuo, Mikko and Luoma, Juha},
  journal={European transport research review},
  volume={9},
  pages={1--8},
  year={2017},
  publisher={Springer}
}

@article{liu2022impact,
  title={Impact of vehicle type, tyre feature and driving behaviour on tyre wear under real-world driving conditions},
  author={Liu, Ye and Chen, Haibo and Wu, Sijin and Gao, Jianbing and Li, Ying and An, Zihao and Mao, Baohua and Tu, Ran and Li, Tiezhu},
  journal={Science of the Total Environment},
  volume={842},
  pages={156950},
  year={2022},
  publisher={Elsevier}
}

@article{chiasserini2017advertisement,
  title={Advertisement delivery and display in vehicular networks: Using v2v communications for targeted ads},
  author={Chiasserini, Carla-Fabiana and Malandrino, Francesco and Sereno, Matteo},
  journal={IEEE Vehicular Technology Magazine},
  volume={12},
  number={3},
  pages={65--72},
  year={2017},
  publisher={IEEE}
}

@article{wang2019efficiently,
  title={Efficiently targeted billboard advertising using crowdsensing vehicle trajectory data},
  author={Wang, Liang and Yu, Zhiwen and Yang, Dingqi and Ma, Huadong and Sheng, Hao},
  journal={IEEE transactions on industrial informatics},
  volume={16},
  number={2},
  pages={1058--1066},
  year={2019},
  publisher={IEEE}
}

@article{einziger2018scheduling,
  title={Scheduling advertisement delivery in vehicular networks},
  author={Einziger, Gil and Chiasserini, Carla Fabiana and Malandrino, Francesco},
  journal={IEEE Transactions on Mobile Computing},
  volume={17},
  number={12},
  pages={2882--2897},
  year={2018},
  publisher={IEEE}
}

@article{abrougui2010location,
  title={Location-aided gateway advertisement and discovery protocol for VANets},
  author={Abrougui, Kaouther and Boukerche, Azzedine and Pazzi, Richard Werner Nelem},
  journal={IEEE Transactions on Vehicular technology},
  volume={59},
  number={8},
  pages={3843--3858},
  year={2010},
  publisher={IEEE}
}

@inproceedings{pradipta2011profiling,
  title={Profiling-based mobile advertisement as a marketing strategy for GPS-based online traffic map},
  author={Pradipta, Sonny and Endarnoto, Sri Krisna and Purnama, James and Nugroho, Anto S and Pawitra, Franciscus C},
  booktitle={Proceedings of the 2011 International Conference on Electrical Engineering and Informatics},
  pages={1--4},
  year={2011},
  organization={IEEE}
}

@article{baecke2017value,
  title={The value of vehicle telematics data in insurance risk selection processes},
  author={Baecke, Philippe and Bocca, Lorenzo},
  journal={Decision Support Systems},
  volume={98},
  pages={69--79},
  year={2017},
  publisher={Elsevier}
}

@article{bian2018good,
  title={Good drivers pay less: A study of usage-based vehicle insurance models},
  author={Bian, Yiyang and Yang, Chen and Zhao, J Leon and Liang, Liang},
  journal={Transportation research part A: policy and practice},
  volume={107},
  pages={20--34},
  year={2018},
  publisher={Elsevier}
}

@article{arumugam2019survey,
  title={A survey on driving behavior analysis in usage based insurance using big data},
  author={Arumugam, Subramanian and Bhargavi, R},
  journal={Journal of Big Data},
  volume={6},
  pages={1--21},
  year={2019},
  publisher={Springer}
}

@article{duri2004data,
  title={Data protection and data sharing in telematics},
  author={Duri, Sastry and Elliott, Jeffrey and Gruteser, Marco and Liu, Xuan and Moskowitz, Paul and Perez, Ronald and Singh, Moninder and Tang, Jung-Mu},
  journal={Mobile networks and applications},
  volume={9},
  pages={693--701},
  year={2004},
  publisher={Springer}
}

@incollection{van2022informational,
  title={Informational rights, informational wrongs: Regulating connected car data access and use for telematics insurance in Europe},
  author={van den Boom, Freyja},
  booktitle={Law, Regulation and Governance in the Information Society},
  pages={307--320},
  year={2022},
  publisher={Routledge}
}

@inproceedings{purwanto2022customer,
  title={Customer Satisfaction and Revisit Intention Modeling for Dining Restaurants in Surabaya},
  author={Purwanto, Dhimas Aditya Putera and Rahayu, Siti and Megawati, Veny},
  booktitle={19th International Symposium on Management (INSYMA 2022)},
  pages={1032--1038},
  year={2022},
  organization={Atlantis Press}
}

@article{bauer2016location,
  title={Location-based advertising on mobile devices: A literature review and analysis},
  author={Bauer, Christine and Strauss, Christine},
  journal={Management review quarterly},
  volume={66},
  number={3},
  pages={159--194},
  year={2016},
  publisher={Springer}
}

@article{tong2020personalized,
  title={Personalized mobile marketing strategies},
  author={Tong, Siliang and Luo, Xueming and Xu, Bo},
  journal={Journal of the Academy of Marketing Science},
  volume={48},
  pages={64--78},
  year={2020},
  publisher={Springer}
}

@article{toch2012personalization,
  title={Personalization and privacy: a survey of privacy risks and remedies in personalization-based systems},
  author={Toch, Eran and Wang, Yang and Cranor, Lorrie Faith},
  journal={User Modeling and User-Adapted Interaction},
  volume={22},
  pages={203--220},
  year={2012},
  publisher={Springer}
}

@inproceedings{tsugawa2002inter,
  title={Inter-vehicle communications and their applications to intelligent vehicles: an overview},
  author={Tsugawa, Sadayuki},
  booktitle={Intelligent Vehicle Symposium, 2002. IEEE},
  volume={2},
  pages={564--569},
  year={2002},
  organization={IEEE}
}

@inproceedings{demba2018vehicle,
  title={Vehicle-to-vehicle communication technology},
  author={Demba, Albert and M{\"o}ller, Dietmar PF},
  booktitle={2018 IEEE international conference on electro/information technology (EIT)},
  pages={0459--0464},
  year={2018},
  organization={IEEE}
}

@article{ring2015connected,
  title={Connected cars--the next targe tfor hackers},
  author={Ring, Tim},
  journal={Network Security},
  volume={2015},
  number={11},
  pages={11--16},
  year={2015},
  publisher={Elsevier}
}

@inproceedings{jafarnejad2015car,
  title={A car hacking experiment: When connectivity meets vulnerability},
  author={Jafarnejad, Sasan and Codeca, Lara and Bronzi, Walter and Frank, Raphael and Engel, Thomas},
  booktitle={2015 IEEE globecom workshops (GC Wkshps)},
  pages={1--6},
  year={2015},
  organization={IEEE}
}

@article{salhieh2021integrating,
  title={Integrating vehicle tracking and routing systems in retail distribution management},
  author={Salhieh, Loay and Shehadeh, Mohammad and Abushaikha, Ismail and Towers, Neil},
  journal={International Journal of Retail \& Distribution Management},
  volume={49},
  number={8},
  pages={1154--1177},
  year={2021},
  publisher={Emerald Publishing Limited}
}

@inproceedings{liaw2002time,
  title={Time-series field trip data analysis using adaptive recognition approach. Analysis on driving patterns and vehicle usage for electric vehicles},
  author={Liaw, Bor Yann and Bethune, Keith P and Kim, Chul Soo},
  booktitle={Proceedings of the 19th Electric Vehicle Symposium (EVS-19)},
  pages={19--23},
  year={2002}
}

@article{idachaba2011design,
  title={Design of a GPS/GSM based tracker for the location of stolen items and kidnapped or missing persons in Nigeria},
  author={Idachaba, Francis Enejo},
  journal={ARPN Journal of Engineering and Applied Sciences},
  volume={6},
  number={10},
  pages={56--60},
  year={2011}
}

@article{johnson2006vehicle,
  title={Vehicle crime: Communicating spatial and temporal patterns},
  author={Johnson, Shane D and Summers, Luc{\'\i}a and Pease, Ken},
  year={2006},
  publisher={UCL Jill Dando Institute of Crime Science}
}

@book{morgan2016reducing,
  title={Reducing criminal opportunity: vehicle security and vehicle crime},
  author={Morgan, Nick and Shaw, Oliver and Feist, Andy and Byron, Christos},
  year={2016},
  publisher={Home Office}
}

@article{connealy2019risk,
  title={Risk factor and high-risk place variations across different robbery targets in Denver, Colorado},
  author={Connealy, Nathan T and Piza, Eric L},
  journal={Journal of criminal justice},
  volume={60},
  pages={47--56},
  year={2019},
  publisher={Elsevier}
}

@article{stevens2021cyber,
  title={Cyber stalking, cyber harassment, and adult mental health: A systematic review},
  author={Stevens, Francesca and Nurse, Jason RC and Arief, Budi},
  journal={Cyberpsychology, Behavior, and Social Networking},
  volume={24},
  number={6},
  pages={367--376},
  year={2021},
  publisher={Mary Ann Liebert, Inc., publishers 140 Huguenot Street, 3rd Floor New~…}
}

@article{georgiadou2019location,
  title={Location Privacy in the Wake of the GDPR},
  author={Georgiadou, Yola and de By, Rolf A and Kounadi, Ourania},
  journal={ISPRS international journal of geo-information},
  volume={8},
  number={3},
  pages={157},
  year={2019},
  publisher={MDPI}
}

@article{bari2022service,
  title={Service headway distribution analysis of FASTag lanes under mixed traffic conditions},
  author={Bari, Chintaman Santosh and Chandra, Satish and Dhamaniya, Ashish},
  journal={Physica A: Statistical Mechanics and its Applications},
  volume={604},
  pages={127904},
  year={2022},
  publisher={Elsevier}
}

@inproceedings{gunjal2023survey,
  title={A Survey On FAScam: FAStag Fraud Detection System},
  author={Gunjal, Siddhant and Tomy, Tini and Serpes, Alessandra and Yadav, Abhishek and Bodade, Vaishali},
  booktitle={2023 5th Biennial International Conference on Nascent Technologies in Engineering (ICNTE)},
  pages={1--5},
  year={2023},
  organization={IEEE}
}

@inproceedings{thaduri2016context,
  title={Context-based maintenance and repair shop suggestion for a moving vehicle},
  author={Thaduri, Adithya and Galar, Diego and Kumar, Uday and Verma, Ajit Kumar},
  booktitle={Current Trends in Reliability, Availability, Maintainability and Safety: An Industry Perspective},
  pages={67--81},
  year={2016},
  organization={Springer}
}

@article{brockett2007biological,
  title={Biological and psychobehavioral correlates of credit scores and automobile insurance losses: Toward an explication of why credit scoring works},
  author={Brockett, Patrick L and Golden, Linda L},
  journal={Journal of Risk and Insurance},
  volume={74},
  number={1},
  pages={23--63},
  year={2007},
  publisher={Wiley Online Library}
}

@inproceedings{roeschlin2018bionyms,
  title={Bionyms: Driver-centric message authentication using biometric measurements},
  author={Roeschlin, Marc and Vaas, Christian and Rasmussen, Kasper B and Martinovic, Ivan},
  booktitle={2018 IEEE Vehicular Networking Conference (VNC)},
  pages={1--8},
  year={2018},
  organization={IEEE}
}

@article{singh2010voice,
  title={Voice Recognition In Automobiles},
  author={Singh, Sarbjeet and Singh, Sukhvinder and Kour, Mandeep and Manhas, Sonia},
  journal={International Journal of Computer Applications},
  volume={6},
  number={6},
  pages={7--11},
  year={2010},
  publisher={Citeseer}
}

@inproceedings{reininger2015first,
  title={A first look at vehicle data collection via smartphone sensors},
  author={Reininger, Michael and Miller, Seth and Zhuang, Yanyan and Cappos, Justin},
  booktitle={2015 IEEE Sensors Applications Symposium (SAS)},
  pages={1--6},
  year={2015},
  organization={IEEE}
}

@misc{palm2021ethical,
  title={Ethical Hacking of Android Auto in the Context of Road Safety},
  author={Palm, Alexander and Gafvelin, Benjamin},
  year={2021}
}

@misc{qatarprivacy2016,
  author       = {{National Cyber Security Agency}},
  title        = {Personal Data Privacy Protection Law - Qatar},
  url          = {https://assurance.ncsa.gov.qa/en/privacy/law}
}

@article{yuvaraj2018me,
  title={How about me? The scope of personal information under the Australian Privacy Act 1988},
  author={Yuvaraj, Joshua},
  journal={Computer law \& security review},
  volume={34},
  number={1},
  pages={47--66},
  year={2018},
  publisher={Elsevier}
}

@article{taylor2020personal,
  title={“Personal information” and group data under the'Privacy Act 1988'(Cth)},
  author={Taylor, Mark J},
  journal={AUSTRALIAN LAW JOURNAL},
  volume={94},
  number={10},
  pages={730--740},
  year={2020},
  publisher={HeinOnline}
}

@article{calzada2022citizens,
  title={Citizens’ data privacy in china: The state of the art of the personal information protection law (pipl)},
  author={Calzada, Igor},
  journal={Smart Cities},
  volume={5},
  number={3},
  pages={1129--1150},
  year={2022},
  publisher={MDPI}
}

@article{zhou2024understanding,
  title={Understanding Chinese Internet Users' Perceptions of, and Online Platforms' Compliance with, the Personal Information Protection Law (PIPL)},
  author={Zhou, Morgana Mo and Qu, Zhiyan and Wan, Jinhan and Wen, Bo and Yao, Yaxing and Lu, Zhicong},
  journal={Proceedings of the ACM on Human-Computer Interaction},
  volume={8},
  number={CSCW1},
  pages={1--26},
  year={2024},
  publisher={ACM New York, NY, USA}
}

@article{chen2021understanding,
  title={Understanding the chinese data security law},
  author={Chen, Jihong and Sun, Jiabin},
  journal={International Cybersecurity Law Review},
  volume={2},
  number={2},
  pages={209--221},
  year={2021},
  publisher={Springer}
}

@article{chik2013singapore,
  title={The Singapore Personal Data Protection Act and an assessment of future trends in data privacy reform},
  author={Chik, Warren B},
  journal={Computer Law \& Security Review},
  volume={29},
  number={5},
  pages={554--575},
  year={2013},
  publisher={Elsevier}
}

@article{schwanen2023getting,
  title={Getting Personal: The Promise and Potential Missteps of Canada’s New Privacy Legislation},
  author={Schwanen, Daniel},
  journal={CD Howe Institute e-Brief},
  volume={349},
  year={2023}
}

@article{jaar2008canadian,
  title={Canadian privacy law: The personal information protection and electronic documents act (PIPEDA)},
  author={Jaar, Dominic and Zeller, Patrick E},
  journal={Int'l. In-House Counsel J.},
  volume={2},
  pages={1135},
  year={2008},
  publisher={HeinOnline}
}

@article{harding2019understanding,
  title={Understanding the scope and impact of the california consumer privacy act of 2018},
  author={Harding, Elizabeth Liz and Vanto, Jarno J and Clark, Reece and Hannah Ji, L and Ainsworth, Sara C},
  journal={Journal of Data Protection \& Privacy},
  volume={2},
  number={3},
  pages={234--253},
  year={2019},
  publisher={Henry Stewart Publications}
}

@article{bond2012eu,
  title={The EU e-Privacy directive and consent to cookies},
  author={Bond, Robert},
  journal={The Business Lawyer},
  pages={215--223},
  year={2012},
  publisher={JSTOR}
}

@article{gdpr2016general,
  title={General data protection regulation},
  author={GDPR, GDPR},
  journal={Regulation (EU)},
  volume={679},
  year={2016}
}

@article{taeihagh2019governing,
  title={Governing autonomous vehicles: emerging responses for safety, liability, privacy, cybersecurity, and industry risks},
  author={Taeihagh, Araz and Lim, Hazel Si Min},
  journal={Transport reviews},
  volume={39},
  number={1},
  pages={103--128},
  year={2019},
  publisher={Taylor \& Francis}
}

@article{fagnant2015preparing,
  title={Preparing a nation for autonomous vehicles: opportunities, barriers and policy recommendations},
  author={Fagnant, Daniel J and Kockelman, Kara},
  journal={Transportation Research Part A: Policy and Practice},
  volume={77},
  pages={167--181},
  year={2015},
  publisher={Elsevier}
}

@article{collingwood2017privacy,
  title={Privacy implications and liability issues of autonomous vehicles},
  author={Collingwood, Lisa},
  journal={Information \& Communications Technology Law},
  volume={26},
  number={1},
  pages={32--45},
  year={2017},
  publisher={Taylor \& Francis}
}

@inproceedings{he2019overview,
  title={Overview of V2V and V2I wireless communication for cooperative vehicle infrastructure systems},
  author={He, Wenxue and Li, Huafu and Zhi, Xiao and Li, Xinghua and Zhang, Jingyuan and Hou, Qian and Li, Yiyao},
  booktitle={2019 IEEE 4th Advanced Information Technology, Electronic and Automation Control Conference (IAEAC)},
  pages={127--134},
  year={2019},
  organization={IEEE}
}

@article{ignatious2022overview,
  title={An overview of sensors in Autonomous Vehicles},
  author={Ignatious, Henry Alexander and Khan, Manzoor and others},
  journal={Procedia Computer Science},
  volume={198},
  pages={736--741},
  year={2022},
  publisher={Elsevier}
}

@article{kabil2022vehicle,
  title={Vehicle to pedestrian systems: Survey, challenges and recent trends},
  author={Kabil, Ahmad and Rabieh, Khaled and Kaleem, Faisal and Azer, Marianne A},
  journal={IEEE Access},
  volume={10},
  pages={123981--123994},
  year={2022},
  publisher={IEEE}
}

@article{kerber2018data,
  title={Data governance in connected cars: the problem of access to in-vehicle data},
  author={Kerber, Wolfgang},
  journal={J. Intell. Prop. Info. Tech. \& Elec. Com. L.},
  volume={9},
  pages={310},
  year={2018},
  publisher={HeinOnline}
}

@misc{tuxera2021,
  title = {Autonomous cars generate more than 300 TB of data per year},
  author = {Simon Wright},
  year = {2021},
  month = {July},
  url = {https://www.tuxera.com/blog/autonomous-cars-300-tb-of-data-per-year/}
}

@article{zanella2014internet,
  title={Internet of things for smart cities},
  author={Zanella, Andrea and Bui, Nicola and Castellani, Angelo and Vangelista, Lorenzo and Zorzi, Michele},
  journal={IEEE Internet of Things journal},
  volume={1},
  number={1},
  pages={22--32},
  year={2014},
  publisher={Ieee}
}

@article{wang2023data,
  title={Data fusion in infrastructure-augmented autonomous driving system: Why? where? and how?},
  author={Wang, Jianda and Wang, Zhendong and Yu, Bo and Tang, Jie and Song, Shuaiwen Leon and Liu, Cong and Hu, Yang},
  journal={IEEE Internet of Things Journal},
  volume={10},
  number={18},
  pages={15857--15871},
  year={2023},
  publisher={IEEE}
}

@article{chamola2024overtaking,
  title={Overtaking mechanisms based on augmented intelligence for autonomous driving: Datasets, methods, and challenges},
  author={Chamola, Vinay and Chougule, Amit and Sam, Aishwarya and Hussain, Amir and Yu, F Richard},
  journal={IEEE Internet of Things Journal},
  year={2024},
  publisher={IEEE}
}

@article{aledhari2023motion,
  title={Motion comfort optimization for autonomous vehicles: Concepts, methods, and techniques},
  author={Aledhari, Mohammed and Rahouti, Mohamed and Qadir, Junaid and Qolomany, Basheer and Guizani, Mohsen and Al-Fuqaha, Ala},
  journal={IEEE Internet of Things Journal},
  volume={11},
  number={1},
  pages={378--402},
  year={2023},
  publisher={IEEE}
}

@article{alladi2020consumer,
  title={Consumer IoT: Security vulnerability case studies and solutions},
  author={Alladi, Tejasvi and Chamola, Vinay and Sikdar, Biplab and Choo, Kim-Kwang Raymond},
  journal={IEEE Consumer Electronics Magazine},
  volume={9},
  number={2},
  pages={17--25},
  year={2020},
  publisher={IEEE}
}

@article{alladi2022comprehensive,
  title={A comprehensive survey on the applications of blockchain for securing vehicular networks},
  author={Alladi, Tejasvi and Chamola, Vinay and Sahu, Nishad and Venkatesh, Vishnu and Goyal, Adit and Guizani, Mohsen},
  journal={IEEE Communications Surveys \& Tutorials},
  volume={24},
  number={2},
  pages={1212--1239},
  year={2022},
  publisher={IEEE}
}

@article{golda2024privacy,
  title={Privacy and security concerns in generative AI: a comprehensive survey},
  author={Golda, Abenezer and Mekonen, Kidus and Pandey, Amit and Singh, Anushka and Hassija, Vikas and Chamola, Vinay and Sikdar, Biplab},
  journal={IEEE Access},
  year={2024},
  publisher={IEEE}
}

@article{chamola2021information,
  title={Information security in the post quantum era for 5G and beyond networks: Threats to existing cryptography, and post-quantum cryptography},
  author={Chamola, Vinay and Jolfaei, Alireza and Chanana, Vaibhav and Parashari, Prakhar and Hassija, Vikas},
  journal={Computer Communications},
  volume={176},
  pages={99--118},
  year={2021},
  publisher={Elsevier}
}

@article{khan2025decentralized,
  title={A Decentralized, Secure, and Reliable Vehicle Platoon Formation With Privacy Protection for Autonomous Vehicles},
  author={Khan, Rabia and Mehmood, Amjad and Song, Houbing and Maple, Carsten},
  journal={IEEE Transactions on Intelligent Transportation Systems},
  year={2025},
  publisher={IEEE}
}

@article{raja2024intuitive,
  title={Intuitive and Privacy-Preserving Traffic Light Control System for Autonomous Vehicles},
  author={Raja, Gunasekaran and Nkenyereye, Lewis and Srividya, Ponnada and Balachandar, Thilaksurya and Senthivel, Sai Ganesh and Mathew, Libin K and Dev, Kapal},
  journal={IEEE Internet of Things Journal},
  year={2024},
  publisher={IEEE}
}

@article{karabulut2023privacy,
  title={Privacy-preserving authentication scheme for connected autonomous vehicles},
  author={Karabulut-Kurt, Gunes and Nari-Baykal, Kubra and Ozdemir, Enver},
  journal={IEEE Transactions on Intelligent Transportation Systems},
  volume={25},
  number={6},
  pages={4998--5010},
  year={2023},
  publisher={IEEE}
}

@article{chen2024empowering,
  title={Empowering IoT-Based Autonomous Driving via Federated Instruction Tuning With Feature Diversity},
  author={Chen, Jiao and He, Jiayi and Chen, Fangfang and Lv, Zuohong and Tang, Jianhua and Jia, Yunjian},
  journal={IEEE Internet of Things Journal},
  year={2024},
  publisher={IEEE}
}

@article{tang2023matching,
  title={Matching 5G connected vehicles to sensed vehicles for safe cooperative autonomous driving},
  author={Tang, Zuoyin and He, Jianhua and Yang, Kun and Chen, Hsiao-Hwa},
  journal={IEEE Network},
  volume={38},
  number={3},
  pages={227--235},
  year={2023},
  publisher={IEEE}
}

@article{hassija2020blockchain,
  title={A blockchain-based framework for lightweight data sharing and energy trading in V2G network},
  author={Hassija, Vikas and Chamola, Vinay and Garg, Sahil and Krishna, Dara Nanda Gopala and Kaddoum, Georges and Jayakody, Dushantha Nalin K},
  journal={IEEE Transactions on Vehicular Technology},
  volume={69},
  number={6},
  pages={5799--5812},
  year={2020},
  publisher={IEEE}
}

@article{hassija2020secure,
  title={Secure lending: Blockchain and prospect theory-based decentralized credit scoring model},
  author={Hassija, Vikas and Bansal, Gaurang and Chamola, Vinay and Kumar, Neeraj and Guizani, Mohsen},
  journal={IEEE Transactions on Network Science and Engineering},
  volume={7},
  number={4},
  pages={2566--2575},
  year={2020},
  publisher={IEEE}
}

@article{grover2021edge,
  title={Edge computing and deep learning enabled secure multitier network for internet of vehicles},
  author={Grover, Harsh and Alladi, Tejasvi and Chamola, Vinay and Singh, Dheerendra and Choo, Kim-Kwang Raymond},
  journal={IEEE Internet of Things Journal},
  volume={8},
  number={19},
  pages={14787--14796},
  year={2021},
  publisher={IEEE}
}

@article{chamola2020fpga,
  title={FPGA for 5G: Re-configurable hardware for next generation communication},
  author={Chamola, Vinay and Patra, Sambit and Kumar, Neeraj and Guizani, Mohsen},
  journal={IEEE Wireless Communications},
  volume={27},
  number={3},
  pages={140--147},
  year={2020},
  publisher={IEEE}
}

@article{verma2019cb,
  title={CB-CAS: Certificate-based efficient signature scheme with compact aggregation for industrial Internet of Things environment},
  author={Verma, Girraj Kumar and Singh, BB and Kumar, Neeraj and Chamola, Vinay},
  journal={IEEE Internet of Things Journal},
  volume={7},
  number={4},
  pages={2563--2572},
  year={2019},
  publisher={IEEE}
}

\vskip -1\baselineskip plus -1fil
\begin{IEEEbiography}[{\includegraphics[width=1in,height=1.5in,clip,keepaspectratio]{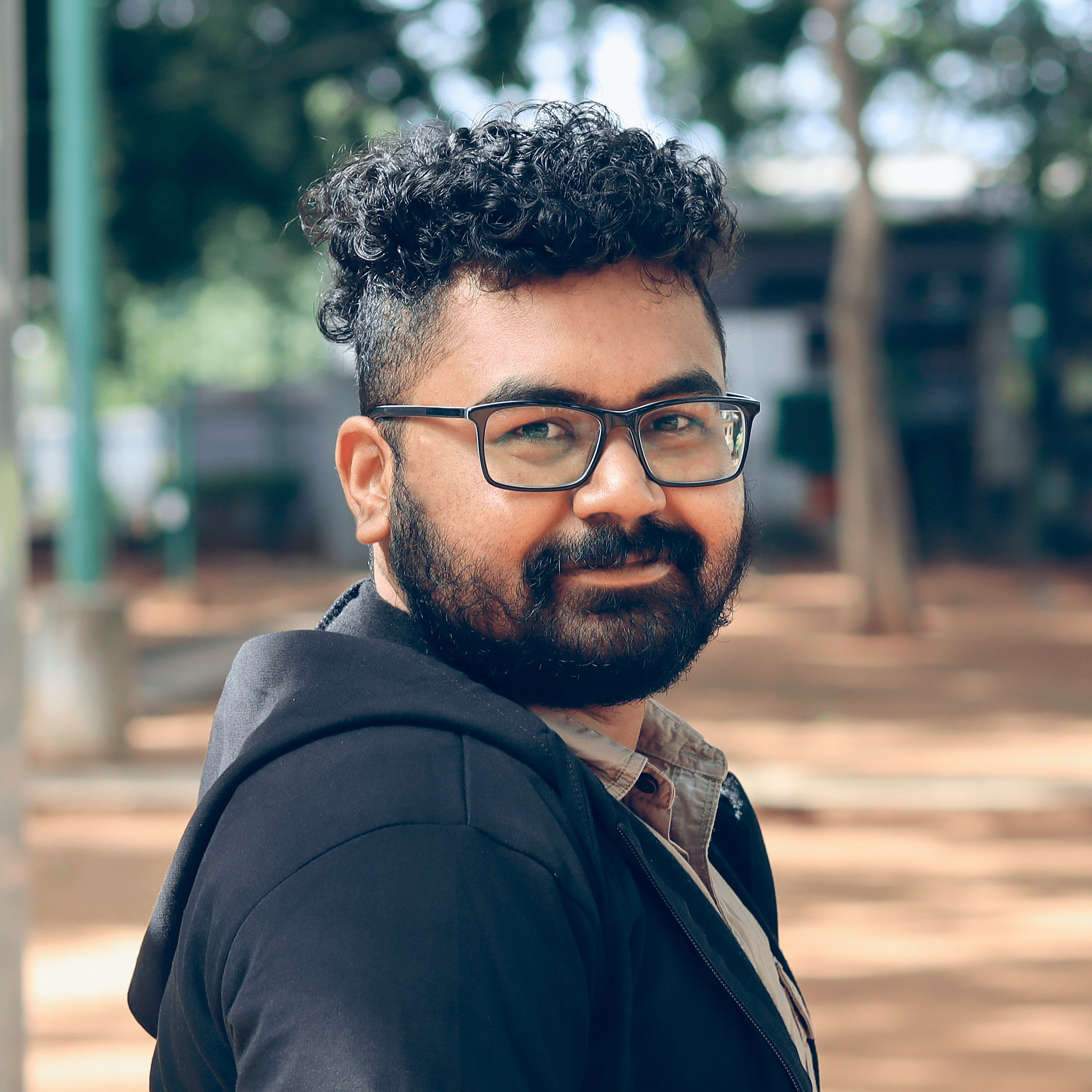}}]{Amit Chougule} holds a PhD in Computer Vision and AI from BITS-Pilani, Pilani Campus, India in 2024. Prior to this, he completed his M.Tech in Big Data and IoT at PES University, Bangalore, India in 2020. In 2023, Chougule contributed as a Visiting Researcher with the Trust in Connected and Autonomous Vehicles (TrustCAV) Research Group at Carleton University, Ottawa, Canada. Throughout his career, he has occupied significant roles as an artificial intelligence and medical imaging researcher within prestigious organizations such as Sony Research, Philips Healthcare, and AIvolved Technologies.
Additionally, Chougule worked as a Senior Research Scientist at Manentia AI \& OptraSCAN, concentrating on medical imaging research. Chougule is a Senior Computer Vision specialist at Lowe’s, where he architects computer vision solutions enabling scalable, data-driven retail transformation.
His research efforts primarily concentrate on advancing artificial intelligence solutions for autonomous driving, healthcare and retail domain leveraging his expertise in computer vision and deep learning methodologies.
\end{IEEEbiography}

\vskip -2\baselineskip plus -1fil
\begin{IEEEbiography}[{\includegraphics[width=1in,height=1.5in,clip,keepaspectratio]{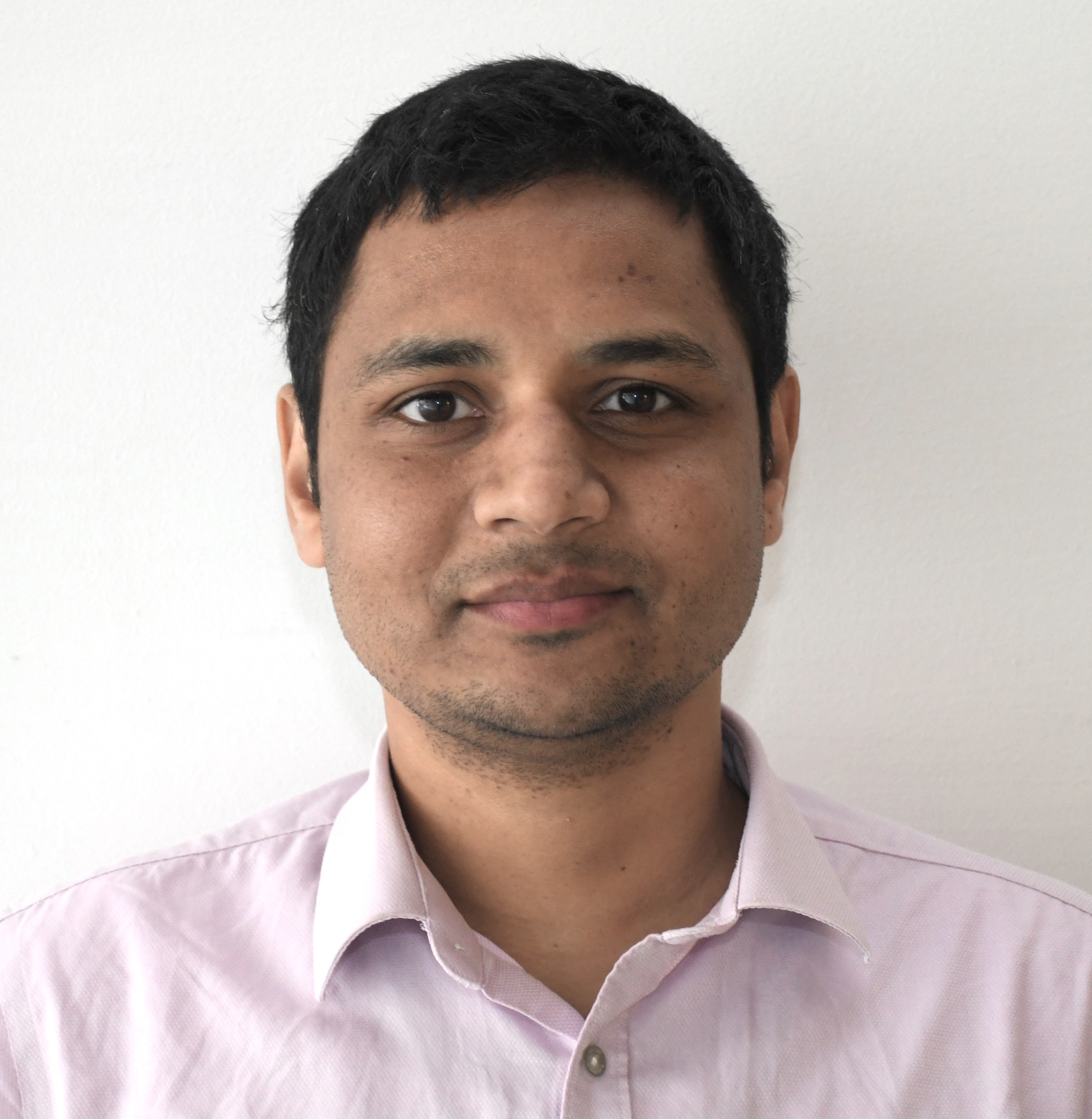}}]{Vinay Chamola} received the B.E. degree in electrical and electronics engineering and master’s degree in communication engineering from the Birla Institute of Technology and Science, Pilani, India, in 2010 and 2013, respectively. He received his Ph.D. in electrical and computer engineering from the National University of Singapore, Singapore 2016. In 2015, he was a Visiting Researcher with the Autonomous Networks Research Group (ANRG), University of Southern California, Los Angeles, CA, USA. He also worked as a post-doctoral research fellow at the National University of Singapore, Singapore. 
He is an Associate Professor with the Department of Electrical and Electronics Engineering, BITS-Pilani, Pilani, where he heads the Internet of Things Research Group / Lab. His research interests include IoT Security, Blockchain, UAVs, VANETs, 5G, and Healthcare. He serves as an Area Editor for the Ad Hoc Networks Journal, Elsevier, and the IEEE Internet of Things Magazine. He is also an Associate Editor in the IEEE Transactions on Intelligent Transportation Systems, IEEE Networking Letters, IEEE Consumer Electronics magazine, IET Quantum Communications, IET Networks, and several other journals. He serves as co-chair of various reputed workshops like IEEE Globecom Workshop 2021, IEEE INFOCOM 2022 workshop, IEEE ANTS 2021, and IEEE ICIAfS 2021, to name a few. He is listed in the World’s Top 2\% Scientists identified by Stanford University. He is the co-founder and President of a healthcare startup, Medsupervision Pvt. Ltd. He is a senior member of the IEEE and a Fellow of the IET.
\end{IEEEbiography}

\vskip -2\baselineskip plus -1fil
\begin{IEEEbiography}[{\includegraphics[width=1in,height=1.5in,clip,keepaspectratio]{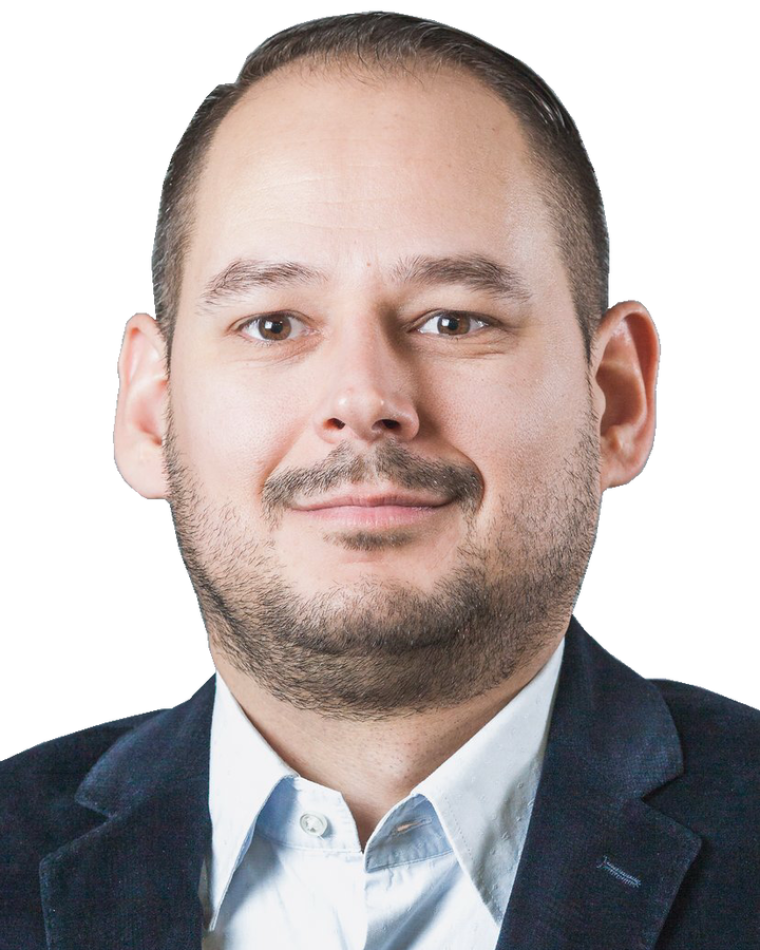}}]{Norbert Herencsar} received the Ph.D. degree from the Brno University of Technology (BUT), Czechia, in 2010. Since 2015, he is an Assoc. Prof. with the Dept. of Telecommunications, BUT. He has authored 140 peer-reviewed journal and 123 conference proceedings articles. His research interests include electronics, fractional-order systems, IoT, and sensors. He serves as the Editor-in-Chief of \textit{IEEE Consumer Electronics Magazine} and as an Associate Editor for \textit{IEEE Transactions on Circuits and Systems II: Express Briefs} journals.
\end{IEEEbiography}

\vskip -1\baselineskip plus -1fil
\begin{IEEEbiography}[{\includegraphics[width=1in,height=1.25in,clip,keepaspectratio]{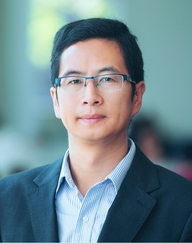}}]{Fei Richard Yu} received the PhD degree in electrical engineering from the University of British Columbia (UBC) in 2003. From 2002 to 2006, he was with Ericsson (in Lund, Sweden) and a start-up (in San Diego, CA, USA), where he worked on the research and development in the areas of advanced wireless communication technologies and new standards. He joined Carleton School of Information Technology and the Department of Systems and Computer Engineering (cross-appointment) at Carleton University, Ottawa, in 2007, where he is currently a Professor. His research interests include cyber-security, connected and autonomous vehicles, artificial intelligence, blockchain, and wireless systems. He has published 600+ papers in reputable journals/conferences, 8 books, and 28 granted patents, with 10,000+ citations (Google Scholar). He received the IEEE TCGCC Best Journal Paper Award in 2019, Distinguished Service Awards in 2019 and 2016, Outstanding Leadership Award in 2013, Carleton Research Achievement Awards in 2012 and 2021, the Ontario Early Researcher Award (formerly Premiers Research Excellence Award) in 2011, the Excellent Contribution Award at IEEE/IFIP TrustCom 2010, the Leadership Opportunity Fund Award from Canada Foundation of Innovation in 2009 and the Best Paper Awards at IEEE ICNC 2018, VTC 2017 Spring, ICC 2014, Globecom 2012, IEEE/IFIP TrustCom 2009 and Int'l Conference on Networking 2005.
He serves on the editorial boards of several journals, including Co-Editor-in-Chief for Ad Hoc \& Sensor Wireless Networks, Lead Series Editor for IEEE Transactions on Vehicular Technology, IEEE Communications Surveys \& Tutorials, and IEEE Transactions on Green Communications and Networking. He has served as the Technical Program Committee (TPC) Co-Chair of numerous conferences. He is a ``Highly Cited Researcher" according to Web of Science since 2019. He is an IEEE Distinguished Lecturer of both Vehicular Technology Society (VTS) and Comm. Society. He is an elected member of the Board of Governors of the IEEE VTS and Editor-in-Chief for IEEE VTS Mobile World newsletter. He is a Fellow of the IEEE, Canadian Academy of Engineering (CAE), Engineering Institute of Canada (EIC), and Institution of Engineering and Technology (IET).

\end{IEEEbiography}

\end{document}